\documentclass[12pt]{article}

\usepackage{tabularx}
\usepackage[DIV13]{typearea}
\usepackage{indentfirst}
\usepackage{amsmath}
\usepackage{amsfonts}
\usepackage{mathrsfs}
\usepackage{amssymb}
\usepackage{mathtools}
\usepackage{bbm}
\usepackage{epsfig}
\usepackage{graphicx}
\usepackage{subfigure}
\usepackage{slashed}
\usepackage{multicol}
\usepackage[usenames,dvipsnames]{color}
\usepackage{cite}
\RequirePackage[colorlinks=true,urlcolor=blue,anchorcolor=blue,citecolor=blue,filecolor=blue,
               linkcolor=blue,menucolor=blue,linktocpage=true,pdfproducer=medialab]{hyperref}
\usepackage{cancel}
\usepackage{feynmp}
\usepackage{lscape}
\usepackage{multirow}
\usepackage{array}
\usepackage{pdfpages}
\usepackage{fancyhdr}
\usepackage{pifont}

\usepackage[left=.9in, right=.9in]{geometry}

\newcommand{\email}[1]{\href{mailto:#1}{\tt #1}}

\numberwithin{equation}{section}
\newcommand{\LL}{\mathscr{L}}

\def\cF{{\cal F}}

\def\cP{{\cal P}}

\def\be{\begin{equation}}
\def\ee{\end{equation}}
\def\beq{\begin{equation}}
\def\eeq{\end{equation}}
\def\bc{\begin{center}}
\def\ec{\end{center}}
\def\bea{\begin{eqnarray}}
\def\eea{\end{eqnarray}}

\def\nn{\nonumber}

\renewcommand{\a}{\alpha}
\renewcommand{\b}{\beta}
\renewcommand{\d}{\delta}
\newcommand{\g}{\gamma}

\newcommand{\ppi}{\mathbf{\pi}}

\newcommand{\UH}{\mathbf{U}}

\newcommand{\VL}{\mathbf{V}}

\newcommand{\DLR}{\mathbf{D}}

\newcommand{\Tr}{{\rm Tr}}

\renewcommand{\to}{\rightarrow}

\newcommand{\dmu}{\partial_\mu}

\usepackage{feynmp,graphicx}
\usepackage[english]{babel}
\usepackage[utf8x]{inputenc}
\usepackage{amsmath,amssymb,lmodern}
\usepackage{graphicx}
\usepackage{bm}
\usepackage{multicol}
\usepackage{xcolor}

\renewcommand{\d}{\delta}
\newcommand{\lam}{\lambda}
\renewcommand{\P}{P}
\renewcommand{\O}{\mathcal{O}}
\renewcommand{\div}{\Delta_\varepsilon}
\newcommand{\der}{\partial}
\renewcommand{\L}{\mathcal{L}}
\newcommand{\F} {\mathcal{F}}

\newcommand{\D} {\mathcal{D}}
\renewcommand{\P} {\mathcal{P}}
\newcommand{\U} {\textbf{U}}
\newcommand{\V} {\textbf{V}}

\newcommand{\bpi} {\bm{\pi}}
\newcommand{\btau} {\bm{\tau}}

\newcommand{\De} { \Delta _{\varepsilon }}

\let\vev\VEV
\def\ac{a_C}
\def\bc{b_C}
\def\ah{a_H}
\def\bh{b_H}
\def\a{\alpha}
\def\b{\beta}
\def\g{\gamma}
\def\Tr{{\rm Tr}}

\def\beq{\begin{equation}}
\def\eeq{\end{equation}}
\def\eps{\varepsilon}

\def \bt#1 {\blue{\bm{#1}}}
\newcommand{\ba} {\begin{equation}\begin{aligned}}
\newcommand{\ea} {\end{aligned}\end{equation}}
\def\bea{\begin{eqnarray}}
\def\eea{\end{eqnarray}}

\newcommand{\blue}[1]{\color{blue} \textbf{#1} \color{black}}

\def\l{\left(}
\def\r{\right)}

 \def\mudmu{16\pi^2\frac{d}{d \ln \mu}}
\newcommand{\paren}[1]{\left( #1 \right)}
\allowdisplaybreaks

\begin{document}
\begin{titlepage}
\vspace*{-1cm}
\phantom{hep-ph/***} 
{\flushleft
{\blue{FTUAM-14-29}}
\hfill{\blue{IFT-UAM/CSIC-14-071}}}\\
\hfill{\blue{DFPD2014/TH/15}}\\
\vskip 1cm
\begin{center}
\mathversion{bold}
{\LARGE\bf On the renormalization of the electroweak\\ 
\vspace{0.3cm}
chiral Lagrangian with a Higgs}
\mathversion{normal}
\vskip .3cm
\end{center}
\vskip 0.5  cm
\begin{center}
{\large M.B.~Gavela}~$^{a)}$
{\large K.~Kanshin}~$^{b)}$,
{\large P.~A.~N.~Machado}~$^{a)}$,
{\large S.~Saa}~$^{a)}$
\\
\vskip .7cm
{\footnotesize
$^{a)}$~
Departamento de F\'isica Te\'orica and Instituto de F\'{\i}sica Te\'orica, IFT-UAM/CSIC,\\
Universidad Aut\'onoma de Madrid, Cantoblanco, 28049, Madrid, Spain\\
\vskip .1cm
$^{b)}$~
Dipartimento di Fisica e Astronomia ``G.~Galilei'', Universit\`a di Padova and \\
INFN, Sezione di Padova, Via Marzolo~8, I-35131 Padua, Italy
\vskip .3cm
\begin{minipage}[l]{.9\textwidth}
\begin{center} 
\textit{E-mail:} 
\email{belen.gavela@uam.es}
\email{kanshin@pd.infn.it,}
\email{sara.saa@uam.es},
\email{pedro.machado@uam.es}
\end{center}
\end{minipage}
}
\end{center}
\vskip 0.5cm

\begin{abstract}

We consider the scalar sector of the effective non-linear electroweak
Lagrangian with a light ``Higgs" particle, up to four derivatives in
the chiral expansion.  The complete off-shell renormalization
procedure is implemented, including one loop corrections stemming from
the leading two-derivative terms, for finite Higgs mass. This
determines the complete set of independent chiral invariant scalar
counterterms required for consistency; these include bosonic operators
often disregarded. Furthermore, new counterterms involving the Higgs
particle which are apparently chiral non -invariant are identified in
the perturbative analysis.  A novel general parametrization of the
pseudoescalar field redefinitions is proposed, which reduces to the
various usual ones for specific values of its parameter; the non-local
field redefinitions reabsorbing all chiral non-invariant counterterms are
then explicitly determined. The physical results translate into renormalization
group equations which may be useful when comparing future Higgs data
at different energies.

  \end{abstract}
\end{titlepage}
\setcounter{footnote}{0}

\pdfbookmark[1]{Table of Contents}{tableofcontents}
\tableofcontents

\newpage

\section{Introduction}
\label{sec:intro}
The field of particle physics is at a most interesting cross-roads, in
which the fantastic discovery of a light Higgs particle has not been
accompanied up to now by any sign of new exotic resonances.  If the
situation persists, either the so-called electroweak hierarchy problem
should stop being considered a problem, with the subsequent revolution
and abandon of the historically successful paradigm that fine-tunings
call for physical explanations -recall for instance the road to the
prediction and discovery of the charm particle, or a questioning of
widespread expectations about the nature of physics at the TeV is
called for.

Indeed, the experimental lack of resonances other than the Higgs
particle casts serious questions on the most popular beyond the
Standard Model (BSM) scenarios devised to confront the electroweak
hierarchy problem, such as low-energy supersymmetry. While there is
still much space for the latter to appear in data to come, it is
becoming increasingly pertinent to explore an alternative solution:
the possibility that the lightness of the Higgs is due to its being a
pseudo-goldstone boson of some strongly interacting physics, whose
scale would be higher than the electroweak one. After all, all
previously known pseudoscalar particles are understood as goldstone or
pseudo-goldstone bosons, as for instance the pion and the other scalar
mesons, or the longitudinal components of the $W$ and $Z$ gauge
bosons.

A light Higgs as a pseudo-goldstone boson was proposed already in the
80's~\cite{Kaplan:1983fs, Georgi:1984af}. The initial models assumed a
strong dynamics corresponding to global symmetry groups such as
$SU(5)$ with a characteristic scale $\Lambda_s$.  One of the goldstone bosons
generated upon spontaneous breakdown of that symmetry was identified with 
Higgs particle $h$, with a goldstone boson scale $f$ such that
$\Lambda_s\le 4\pi f$.  The non-zero Higgs mass would result instead from an
explicit breaking of the global symmetry at a lower scale, which
breaks the electroweak symmetry and generates dynamically a potential
 for the Higgs particle~\cite{Dugan:1984hq}.  The
electroweak scale $v$, defined from the $W$ gauge boson mass
$m_W=gv/2$, does not need to coincide neither with the vacuum
expectation value (vev) of the Higgs particle, nor with $f$, although
a relation links them together.  In these hybrid linear/non-linear
constructions, a linear regime is recovered in the limit in which
$\Lambda_s $ - and thus $f$ - goes to infinity.
 
The most successful modern variants of the same idea include $SO(5)$
as strong group~\cite{Agashe:2004rs, Contino:2006qr}, with the nice
new feature that the Standard Model (SM) electroweak interactions
themselves may suffice as agents of the explicit breaking.  This
avenue is being intensively explored, albeit significant fine-tunings
in the fermionic sector~\cite{Panico:2012uw} plague the models
considered up to now.

A model-independent way to approach the low-energy impact of a
pseudo-goldstone nature of the Higgs particle is to use the effective
Lagrangian for a non-linear realisation of electroweak symmetry
breaking (EWSB), as it befits the subjacent strong dynamics. While
decades ago that effective Lagrangian was determined for the case of a
heavy Higgs (that is, a Higgs absent from the low-energy spectrum),
only in recent years the formulation has been extended to include a
light Higgs particle $h$~\cite{Grinstein:2007iv, Contino:2010mh,
  Azatov:2012bz, Alonso:2012px, Brivio:2013pma}. The major differences
are: i) the substitution of the typical functional $h$ dependence in
powers of $(v+h)$ (which holds for the SM and for BSM scenarios with
linearly realised EWSB) by a generic functional dependence on $h/f$;
ii) an operator basis which in all generality differs from that in
linear realizations.

This last point was recently clarified~\cite{ABGMS}. If the
pseudo-goldstone boson $h$ is embedded in the high-energy strong
dynamics as an electroweak doublet, the number of independent
operators coincides with that in linear expansions, as does the
relative weight of gauge couplings for fixed number of external $h$
legs. If instead $h$ was born as a goldstone boson but it was not
embedded in the strong dynamics as an electroweak doublet (e.g. if it is a SM
singlet), the total number of operators is still as in the linear case
but the operators are different: the relative weight of
phenomenological gauge couplings, for a fixed number of external $h$
legs, differs from that in the SM and in linear expansions. The best analysis tool
then is the general non-linear effective Lagrangian, supplemented by
model-dependent relations.  Finally, $h$ may not be a pseudo-goldstone
boson but a generic SM scalar singlet: e.g.  a SM ``impostor", a
dilaton or any dark sector scalar singlet; the appropiate tool then is
that of the non-linear effective Lagrangian with a light $h$ and
completely arbitrary coefficients.  Note that this Lagrangian can in
fact describe all cases mentioned, including the SM one, by setting
constraints on its parameters appropiate to each case, and we will
thus analyze it here in full generality.

More precisely, we will focus on the scalar sector of the non-linear
Lagrangian (i.e. longitudinal
components of the $W$ and $Z$ bosons plus $h$), up to four derivatives in the chiral expansion.  While
previous literature~\footnote{The on-shell precursory study in
  Ref.~\cite{Longhitano:1980tm} assumed no Higgs in the low-energy
  spectrum.},\cite{Delgado:2013hxa, Espriu:2013fia, Delgado:2014jda}
has restrained the one-loop renormalization study of this sector to
on-shell analysis, the complete off-shell renormalization procedure is
implemented in this paper, by considering the one-loop corrections to
the leading - up to two-derivatives - scalar Lagrangian, and
furthermore taking into account the finite Higgs mass.  The off-shell
procedure will allow:
\begin{itemize}
\item To guarantee that all counterterms required for consistency are
  identified, and that the corresponding basis of chirally invariant
  scalar operators is thus complete. It will follow that some
  operators often disregarded previously are mandatory when analysing
  the bosonic sector by itself.
\item To shed light on the expected size of the counterterm
  coefficients, in relation with current controversies on the
  application of ``naive dimensional analysis"
  (NDA)~\cite{Manohar:1983md, Jenkins:2013sda} for light $h$.
\item To identify the renormalization group equations (RGE) for the
  bosonic sector of the chiral Lagrangian.
\end{itemize}

A complete one-loop off-shell renormalization of the electroweak
chiral Lagrangian with a decoupled Higgs particle was performed in the
seminal papers in Ref.~\cite{Appelquist:1980ae}.  Using the non-linear
sigma model and a perturbative analysis, apparently chiral
non-invariant divergences (NID) were shown to appear as counterterms
of four-point functions for the ``pion" fields, in other words, for
the longitudinal components of the $W$ and $Z$ bosons. Physical
consistency was guaranteed as those NID were shown to vanish on-shell
and thus did not contribute to physical amplitudes. They were an
artefact of the perturbative procedure -- which is not explicitly
chiral invariant -- and a redefinition of the pion fields leading to
their reabsortion was identified, see also
Refs.~\cite{Gerstein:1971fm, Weinberg:1968de, Charap:1970xj}.  In the
present work, additional new NID in three and four-point functions
involving the Higgs field will be shown to be present, and their
reabsortion explored.  Furthermore, a general parametrization of the
pseudo-goldstone boson matrix will be formulated, defining a parameter
$\eta$ which reduces to the various usual pion parametrizations for different values of $\eta$, and
the non-physical character of all NID will be analysed.

The resulting RGE restricted to the bosonic sector may eventually
illuminate future experimental searches when comparing data to be
obtained at different energy scales.  The structure of the paper can
be easily inferred from the Table of Contents.

\section{The Lagrangian}
\label{sec:lag}
We will adopt the formulation in Refs.~\cite{Alonso:2012px,
  Agashe:2004rs, Contino:2010mh, Azatov:2012bz, Brivio:2013pma} to
describe in all generality a light scalar boson $h$ in the context of
a generic non-linear realisation of EWSB.  The Lagrangian describes
$h$ as a SM singlet scalar whose couplings do not need to match those
of an $SU(2)$ doublet. The focus of the present analysis will be set
on the physics of the longitudinal components of the gauge bosons
(denoted below as ``pions" $\pi$) and of the $h$ scalar, and only
these degrees of freedom will be explicited below. The corresponding
Lagrangian can be decomposed as
\beq  \label{L}
\L=\L_0 + \L_2 + \L_4\,,
\eeq
where the $\L_i$ subindex indicates number of derivatives: 
\begin{align}
\L_0 =& - V(h)\,\label{L0},
\\[3mm]
\L_{2} =& 
\frac{1}{2} \der_\mu h \der^\mu h \ \F_H(h)
-\frac{v^2}{4} \Tr[\V_\mu  \V^\mu ] \ \F_C(h) \label{L2} \,, 
\\[3mm]
\L_4=&\sum_i c_i \P_i\,.
\label{L4}
\end{align}
In  Eq.~(\ref{L2}) we have omitted the two-derivative custodial breaking operator, because  the size of its coefficient is phenomenologically very strongly constrained. 
 In consequence, and as neither gauge nor Yukawa interactions are considered in this work, no custodial-breaking countertem will be required by the renormalization procedure to be present among the four-derivative operators in $\L_4$. Our analysis is thus restricted to the custodial-preserving sector. 

The $\P_i$ operators in Eq.~(\ref{L4}) are shown explicitly in
Table~\ref{table:L4-terms}, with $c_i$ being arbitrary constant
coefficients; in the SM limit only $a_C$ and $b_C$ would survive, with
$a_C=b_C=1$.
 $V(h)$  in Eq.~(\ref{L0}) denotes a general potential for the $h$ field, for which only up to terms quartic in $h$ will be made explicit, with arbitrary coefficients $\mu_i$ and $\lambda$,
\begin{equation} \label{V}
V \equiv 
  \mu_1^3 \,h + \frac{1}{2} m_h^{2} h^{2} + 
 \dfrac{\mu_3}{3!}  h^{3} +\dfrac{\lam}{4!} h^{4}\,.
 \end{equation}
It will be assumed that the $h$ field is the physical one, with
$\vev{h}=0$: the first term in $V(h)$ is provisionally kept in order to cancel the
tadpole amplitude at one loop; we will clarify this point in
Sect.~\ref{sec:1pt}.

In the expressions above, $\F_i(h)$ are assumed to be generic
polynomials in $h$. The present analysis will only require up to four-field vertices, for which it suffices to explicit the $h$ dependence of
those functions up to quadratic terms. $\F_{H,C}(h)$ will be thus 
parametrized as~\cite{Contino:2010mh} 
\begin{equation}
\F_{H,C}(h)
  \equiv 1 + 2a_{H,C}h/v + b_{H,C}h^2/v^2\,,
  \label{FHC}
  \end{equation}
  while for all $\P_i(h)$ operators in Table~\ref{table:L4-terms} the corresponding functions will be defined as~\footnote{The notation 
differs slightly  from that in
  Ref.~\cite{Brivio:2013pma}: for simplicity, 
  redundant parameters have been eliminated via the replacements
  $\der_\mu\F_i(h)\to\der_\mu h\, \F_i(h)$,
  $\der_\mu\F_i(h)\der_\nu\F'_i(h)\to\der_\mu h \der_\nu h\,
  \F_i(h)$, and $\Box\F_i(h)\to\Box h\,\F_i(h)$.} 
  \begin{equation}
   c_i\F_i(h) \equiv c_i + 2a_ih/v
  + b_ih^2/v^2\,.\label{Fi}
  \end{equation}
Note that in these parametrizations the natural dependence on $h/f$
expected from the underlying models has been traded by $h/v$: the
relative $\xi\equiv v/f<1$ normalization is thus implicitly reabsorbed
in the definition of the constant coefficients, which are then
expected to be small parameters, justifying the truncated
expansion. The case of $\F_{C} (h)$ is special in that the $v^2$
dependence in front of the corresponding term in the Lagrangian
implies a well known fine-tuning to obtain the correct $M_W$ mass,
with $a_C=b_C=1$ in the SM limit. Furthermore, while present data set
strong constraints on departures from SM expectations for the latter,
$a_H$ and $b_H$ could still be large.  Note as well that $v/f$ is not
by itself a physical observable from the point of view of the
low-energy effective Lagrangian.
 
A further comment on $\cF_H(h)$ may be useful:  through a redefinition
of the $h$ field~\cite{Giudice:2007fh} it would be possible to absorb
it completely. Nevertheless, this redefinition would affect all other
couplings in which $h$ participates and induce for instance
corrections on fermionic couplings which are weighted by SM Yukawa
couplings; it is thus pertinent not to disregard $\cF_H(h)$ here, as
otherwise consistency would require to include in the analysis the
corresponding $\cF_i(h)$ fermionic and gauge functions. If a complete
basis including all SM fields is considered assigning individual
arbitrary functions $\cF_i(h)$ to all operators, it would then be possible to redefine
away  completely one $\cF_i(h)$ without loss of generality: it is up to
the practitioner to decide which set of independent operators he/she
may prefer, and to redefine away one of the functions, for instance
$\cF_H(h)$. For the time being, we keep explicit $\cF_H(h)$ all
through, for the sake of generality~\footnote{ Note that $\cF_H(h)$ is
  not expected to be generated from the most popular composite Higgs
  models, as the latter break explicitly the chiral symmetry only via a
  potential for $h$ externally generated, while $\cF_H(h)$ would
  require derivative sources of explicit breaking of the chiral symmetry. A similar comment
  could be applied to $\cP_{\Delta H}$.}.
 
In Eqs~(\ref{L2}) and (\ref{L4}) \mbox{$\VL_\mu\equiv
  \left(\DLR_\mu\UH\right)\UH^\dagger$}, with $\UH(x)$ being the
customary dimensionless unitary matrix describing the longitudinal
degrees of freedom of the three electroweak gauge bosons, which transforms
under the accidental $SU(2)_{L}\times SU(2)_{R}$ global symmetry of the SM scalar
sector as
\beq 
\UH(x) \rightarrow L\, \UH(x) \, R^\dagger\,, 
\label{U}
\eeq 
where $L$, $R$ denote the corresponding $SU(2)_{L,R}$ transformations. Upon EWSB
this symmetry is spontaneoulsy broken to the vector
subgroup. $\VL_\mu$ is thus a vector chiral field belonging to the
adjoint of the global $SU(2)_L$ symmetry. The covariant derivative can
be taken in what follows as given by its pure kinetic term
$\D_\mu=\partial_\mu$, since the transverse gauge field components
will not play a role in this paper.  

 We analyse next the freedom in defining the
$\U$ matrix and work with a general parametri\-zation truncated up to
some order in $\pi/v$. 
On-shell quantities must be independent of the choice of
parametrization for the $\U$ matrix~\cite{Weinberg:1968de}, while it
will be shown below that all NID depend instead on the specific
parametrization chosen. The NID in which the $h$ particle participates
will turn out to offer a larger freedom to be redefined away than the
pure pionic ones.

\subsection{The Lagrangian in a general $\U$ parametrization}
\label{sec:general-u-matrix}
The nonlinear $\sigma$ model can be written as~\cite{Weinberg:1968de}
\begin{equation}
  \L_{\rm NL} = \frac{1}{2}D_\mu\bpi D^\mu\bpi=\frac{v^2}{4}\Tr[\der_\mu\U\der^\mu\U^\dag]
             = \frac{1}{2}G_{ij}(\bpi^2)\der_\mu\pi_i \der^\mu\pi_j\,,
             \label{LNL}
\end{equation}
where $D_\mu$ is a derivative ``covariant'' under the non-linear
chiral symmetry, $\U$ has been defined in Eq.~(\ref{U}) and
$\bpi=(\pi_1,\pi_2,\pi_3)$ represents the pion vector.  In geometric
language, $G_{ij}(\bpi^2)$ can be interpreted as the metric of a
3-sphere in which the pions live, and the freedom of parametrization
is just a coordinate transformation (see Ref.~\cite{Appelquist:1980ae}
and references therein). Indeed, Weinberg has
shown~\cite{Weinberg:1968de} that different linear realizations of the
chiral symmetry would lead to different metrics, which turn out to
correspond to different $\U$ parametrizations; they are all equivalent
with respect to the dynamics of the pion fields as the non-linear
transformation induced on them is unique, and they are connected via
redefinitions of the pion fields. In order to illustrate this
correspondence explicitly, let us define general $X$ and $Y$ functions
as follows:
\begin{equation}
  \U \equiv X(z) + \frac{i \btau \cdot\bpi}{v} Y(z), \qquad z = \bpi^2/v^2 \,,
  \end{equation}
 where $\btau$ denotes the Pauli matrices, and $v$ is the characteristic scale of the $\bpi$ goldstone bosons.
 $X(z)$ and $Y(z)$ are related via the unitarity condition $\U\U^\dag=\bold{1}$, 
  \begin{equation}
    X(z) = \sqrt{1-z Y(z)^2}\,.
    \end{equation}
 The $ G_{ij}$ metric can now be rewritten as 
      \begin{equation}
  G_{ij}(\bpi^2) = Y(z)^{2} \d_{ij} + 4 \l X'(z)^2+z Y'(z)^2+Y(z) Y'(z) \r \frac{\pi_i \pi_j}{v^2},
\end{equation}
where the primes indicate derivatives with respect the the $z$ variable, and $Y(0)=\pm 1$ is   required 
for canonically normalized pion kinetic terms. 

The Lagrangian in Eq.~(\ref{LNL}) is invariant under the
transformation $Y\to -Y$, or equivalently $ \bpi\to-\bpi$. It is easy
to relate $X$ and $Y$ to the functions in Weinberg's
analysis of chiral symmetry~\footnote{The $f(\bpi^2)$ function defined
  in Ref.~\cite{Weinberg:1968de} is related to $X$ and $Y$ simply by
  $f(x)=X(x)/Y(x)$}. A Taylor expansion of $\U$ up to order
$\bpi^{2N+2}$ bears $N$ free parameters.  {\it A priori} the present
analysis requires to consider in $\LL_2$ terms up to $\O(\bpi^6)$, as
the latter may contribute to 4-point functions joining two of its pion
legs into a loop. Nevertheless, the latter results in null contributions for massless
pions,  and in practice it will suffice to consider inside $\U$
up to terms cubic on the pion fields. We thus define a single
parameter $\eta$\, which encodes all the parametrization dependence,
\begin{equation}
Y(z) \equiv 1 + \eta
\,z + O(z^{2})\,, 
\end{equation} 
resulting in 
\begin{equation}
  \U = 1-\frac{\bpi^2}{2v^2} - \l\eta+\frac{1}{8}\r \frac{\bpi^4}{v^4}
        + \frac{i (\bpi\btau)}{v} \l 1+ \eta \frac{\bpi^2}{v^2} \r + \dots
        \label{Uparam}
\end{equation}
Specific values of $\eta$ can be shown to correspond to different parametrizations up
to terms with four pions, for instance:
\begin{itemize}
\item $\eta=0$ yields the square root parametrization: $\U =
\sqrt{1-\bpi^2/v^2}+i(\bpi\btau)/v$\,,
\item $\eta=-1/6$ yields the exponential one:  
$\U =\exp(i\bpi\cdot\tau/v)$\,.
\end{itemize}

The $\L_2$ Lagrangian can now be written in terms of pion
fields. Using the $\F_i(h)$ expansions in Eqs.~(\ref{FHC}) and
(\ref{Fi}) it results
\begin{align} \label{L0ex}
\L_2 =& 
\frac{1}{2} \der_\mu h \der^\mu h  \l 1+2 a_H \frac{h}{v} + b_H \frac{h^2}{v^2} \r
\\
+&\left\{\frac{1}{2} \der_\mu \bpi \der^\mu \bpi + \frac{(\bpi\der_\mu\bpi)^2}{2v^2}
+ \eta\left[ \frac{\bpi^2(\der_\mu\bpi)^2}{v^2}+2\frac{(\bpi\der_\mu\bpi)^2}{v^2} \right] \right\}
 \l 1+2 a_C \frac{h}{v} + b_C \frac{h^2}{v^2} \r\,,
\nonumber%
\end{align}
where terms containing more than four fields are to be disregarded.  The operators required by the renormalization procedure to be present in $\L_4$ as counterterms will be shown below to correspond to those on the left-hand side of Table~\ref{table:L4-terms}, which were already known to constitute an independent and complete set of bosonic four-derivative operators~\cite{Alonso:2012px, Brivio:2013pma}.  
 The expansion up to four fields of the terms in $\L_4$ -Eq.~(\ref{L4})-  is shown on the right column of 
Table~\ref{table:L4-terms}.

\subsubsection*{Counterterm Lagrangian}

\renewcommand{\arraystretch}{1.9}
\begin{table}
\begin{centering}
\begin{tabular}{|p{1.4cm}| m{4.9cm} |p{8.8cm}|}
\hline \multicolumn{2}{|c|}{$\L_4$ operators} & Expansion in $\pi$ fields\\\hline
$c_6\mathcal{P}_6$& $c_6\left[\Tr(V_\mu V^\mu)\right]^2\,\mathcal{F}_6(h)$ &
$\displaystyle \frac{4c_6}{v^4}(\partial_\mu\bpi\partial^\mu\bpi)^2$ \\\hline
$c_7\mathcal{P}_{7}$& $c_7\Tr(V_\mu V^\mu)\frac{1}{v}\Box h\,\mathcal{F}_{7}(h)$ & %
		$\displaystyle -\frac{2c_{7}}{v^3}\Box h (\partial_\nu\bpi\partial^\nu\bpi) -\frac{4a_{7}}{v^4}(h \Box h) (\partial_\nu\bpi\partial^\nu\bpi)$\\\hline
  $c_8\mathcal{P}_{8}$ &
  $c_8\Tr(V_\mu V_\nu)\frac{1}{v^2}\partial^\mu h \partial^\nu h\mathcal\,{F}_{8}(h)$ & 
  $\displaystyle -2\frac{c_{8}}{v^4}(\partial_\mu h\partial_\nu h) (\partial^\mu\bpi\partial^\nu\bpi)$\\ \hline
  $c_9\mathcal{P}_{9}$ & 
  $c_9\Tr[(\mathcal{D}_\mu V^\mu)^2]\,\mathcal{F}_{9}(h)$
  & $\displaystyle -\frac{2c_{9}}{v^4}\big[v^2(\Box\bpi\Box\bpi) +2\eta\bpi^2(\Box\bpi)^2 +\paren{1+4\eta}(\bpi\Box\bpi)^2$\\
  & & $\displaystyle +8\eta(\bpi\der_\mu\bpi)(\der^\mu\bpi\Box\bpi)+\paren{2+4\eta}(\partial_\mu\bpi)^2(\bpi\Box\bpi)\big]$\\
  & & $\displaystyle -\frac{4a_{9}}{v^3}h(\Box\bpi\Box\bpi)-\frac{2b_{9}}{v^4}h^2(\Box\bpi\Box\bpi)$\\ \hline
  $c_{10}\mathcal{P}_{10}$ &
  $c_{10}\Tr(V_\nu\mathcal{D}_\mu V^\mu)\frac{1}{v}\partial^\nu h \,\mathcal{F}_{10}(h)$ &
  $\displaystyle \frac{-2c_{10}}{v^3}\partial^\nu h(\partial_\nu\bpi\Box\bpi)+\frac{-4a_{10}}{v^4} h\partial^\nu h (\partial_\nu\bpi\Box\bpi)$\\ \hline
  $c_{11}\mathcal{P}_{11}$ & 
  $c_{11}\left[\Tr(V_\mu V_\nu)\right]^2\,\mathcal{F}_{11}(h)$ &
  $\displaystyle \frac{4c_{11}}{v^4}(\partial_\mu\bpi\partial_\nu\bpi)^2$ \\ \hline
  $c_{20}\mathcal{P}_{20}$ &
  $c_{20}\Tr(V_\mu V^\mu)\frac{1}{v^2}\partial_\nu h \partial^\nu h\,\mathcal{F}_{20}(h)$ &
  $\displaystyle -2\frac{c_{20}}{v^4}(\partial_\mu h\partial^\mu h) (\partial_\nu\bpi\partial^\nu\bpi)$\\ \hline
  $c_{\Box H}\mathcal{P}_{\Box H}$ &
  $\displaystyle\frac{1}{v^2} (\Box h\Box h)\,\mathcal{F}_{\Box H}(h)$ &
  $\displaystyle \frac{c_{\Box H}}{v^2}\left(\Box h\Box h\right)+\frac{2a_{\Box H}}{v^3}h\left(\Box h\Box h\right)+\frac{b_{\Box H}}{v^4}h^2\left(\Box h\Box h\right)$\\ \hline
  $c_{\Delta H}\mathcal{P}_{\Delta H}$ &
  $\displaystyle\frac{1}{v^3}(\partial_\mu h\partial^\mu h)\Box h \,\mathcal{F}_{h2}(h)$ &
  $\displaystyle \frac{c_{\Delta H}}{v^3}(\partial_\mu h\partial^\mu h)\Box h +\frac{2a_{\Delta H}}{v^4}(\partial_\mu h\partial^\mu h)h\Box h$\\ \hline
  $c_{DH}\mathcal{P}_{DH}$ &
  $\displaystyle\frac{1}{v^4}(\partial_\mu h\partial^\mu h)^2\,\mathcal{F}_{DH}(h)$ &
  $\displaystyle \frac{c_{DH}}{v^4}(\partial_\mu h\partial^\mu h)^2$\\  \hline
\end{tabular}
\par\end{centering}
\caption{The two columns on the left show the operators required  to be in $\L_4$, Eq.~(\ref{L4}), by the renormalization procedure. The right hand side gives the corresponding explicit expansion in terms of pion and $h$ fields (up to four fields), following the $\U$ expansion in Eq.~(\ref{U}) and the $\mathcal{F}_i(h)$ parametrization in Eq.~(\ref{Fi}).}
\label{table:L4-terms}
\end{table}
\renewcommand{\arraystretch}{1}

It is straightforward to obtain the counterterm Lagrangian via the
usual procedure of writing the bare parameters and field wave
functions in terms of the renormalized ones (details in
Appendix~\ref{sec:AppA}),
\begin{align}
\d\L_0+\d\L_2 =&
\frac{1}{2} \der_\mu h \der^\mu h 
\l \d_h + 2 \d \ah  \frac{h}{v} 
+  \d \bh  \frac{h^2}{v^2} \r 
 -\frac{1}{2} \d m_h^{2} h^{2} 
-\d \mu_1^3 h - \dfrac{\d\mu_3}{3!}  h^3 - \dfrac{\d\lam}{4!}  h^4
\nonumber\\ +&
\frac{1}{2} \der_\mu \bpi \der^\mu \bpi 
\l \d_\pi + 2 \d \ac \frac{h}{v} +\d \bc \frac{h^2}{v^2} \r
\nonumber\\+&
\l \d_\pi-\frac{\d v^2}{v^2} \r\frac{1}{2v^2} 
\l (\bpi \der_\mu \bpi)(\bpi \der^\mu \bpi) 
+ 2\eta \l\bpi^2(\der_\mu\bpi)^2+2(\bpi\der_\mu\bpi)^2\r \r.
\end{align}
$\d\L_4$ is simply given by
$\L_4$ with the replacement $c_i,a_i,b_i\to\d c_i,\d a_i,\d b_i$, apart from operator $\P_{9}$, for which 
\begin{align*}
\d (c_{9}\P_{9}) \to
  -& \frac{2\delta c_{9}}{v^4}\big[ (1+4\eta)(\bpi\Box\bpi)^2 
  + 2(1+2\eta)(\bpi\Box\bpi)(\der_\mu\bpi)^2 \nonumber\\
  \,& \qquad\qquad+ 2\eta\bpi^2(\Box\bpi)^2
  + 8\eta(\Box\bpi\der_\mu\bpi)(\bpi\der^\mu\bpi) \big]
\\-&   
\frac{2}{v^2}\Box\bpi\Box\bpi 
\left[ \l\d c_{9} - \frac{\d v^2}{v^2}\r 
  + \frac{2\d a_{9} h}{v} + \frac{\d b_{9} h^2}{v^2}  \right].
\end{align*}
Among the Lagrangian parameters above, $v$ plays the special role of
being the characteristic scale of the goldstone bosons (that is, of
the longitudinal degrees of freedom of the electroweak bosons),
analogous to the pion decay constant in QCD. It turns out that
the counterterm coefficient $\delta v^2=0$ as shown below.
We have left explicit the $\delta v^2$ dependence all
through the paper, though, in case it may be interesting to apply our
results to some scenario which includes sources of explicit chiral
symmetry breaking in a context different than the SM one; it also
serves as a check-point of our computations.

\section{Renormalization of  off-shell Green functions}
\label{sec:renorm}
We present in this section the results for the renormalization of the
1- 2-, 3-, and 4-point functions involving $h$ and/or $\bpi$ in a
general $\U$ parametrization, specified by the $\eta$ parameter in
Eq.~(\ref{Uparam}).  Dimensional regularization is a convenient
regularization scheme as it avoids quadratic divergencies, some of
which would appear to be chiral noninvariant, leading to further
technical complications~\cite{Charap:1970xj, Gerstein:1971fm}. 
Dimensional regularization is thus used below, as well as minimal subtraction scheme
as renormalization procedure. The notation
\begin{equation*}
\De= +\frac{1}{16 \pi^2} \frac{2}{\varepsilon}\,
\end{equation*}
will be adopted, while  FeynRules, FeynArts, and
FormCalc~\cite{Mertig:1990an, Alloul:2013bka, Kublbeck:1990xc, Hahn:2000kx,
  Hahn:1998yk} will be used to compute one-loop amplitudes. Diagrams with closed pion loops give zero contribution for the case of massless pions under study, and any reference to them will be omitted below.

 Table~\ref{table:ct-process} provides and overview of which $\L_4$
 operator coefficients contribute to amplitudes involving pions and/or
 $h$, up to 4-point vertices. It also serves as an advance over the
 results: all operators in (\ref{L4}) will be shown to be required by
 the renormalization procedure. Furthermore we have checked that they
 are all independent and thus their ensemble, when considered by
 itself, constitutes a complete and independent basis of scalar
 operators, up to four-derivatives in the non-linear expansion. None
 of them should be disregarded on arguments of their tradability for
 other operators, for instance for fermionic ones via the application of the equations of motion (EOM), unless the
 latter operators are explicitly included as part of the analysis, or
 without further assumptions (i.e. to neglect all fermion masses). See
 Sect.~\ref{sec:comparison} for comparison with previous literature.
\begin{table}
\begin{centering}
\begin{tabular}{|c|c|c|c|c|c|c|c|}
\hline& \multicolumn{7}{c|}{ Amplitudes}
   \\ \hline
   & $2h$ & $3h$ & $4h$ & $2\pi$ & $2\pi h$ & $2\pi2h$ & $4\pi$  
   \\ \hline
 $\P_6$ & \text{} & \text{} & \text{} & \text{} & \text{} & \text{} & $c_6$ 
 \\ \hline
 $\P_{7}$ & \text{} \text{} & \text{} & \text{} & \text{} & $c_{7}$ & $a_{7}$ & \text{} 
 \\ \hline
 $\P_{8}$ \text{} & \text{} & \text{} & \text{} & \text{} & \text{} & $c_{8}$ & \text{} 
 \\ \hline
 $\P_{9}$ & \text{} & \text{} & \text{} & $c_{9}$ &$a_{9}$ & $b_{9}$ & $c_{9}$ 
 \\ \hline
 $\P_{10}$ & \text{} & \text{} & \text{} & \text{} & $c_{10}$ & $a_{10}$ & \text{} 
 \\ \hline
 $\P_{11}$ & \text{} & \text{} & \text{} & \text{} & \text{} & \text{} & $c_{11}$ 
 \\ \hline
 $\P_{20}$ & \text{} & \text{} & \text{} & \text{} & \text{} & $c_{20}$ & \text{} 
 \\ \hline
 $\P_{\Box H}$ & $c_{\Box H}$ & $a_{\Box H}$ & $b_{\Box H}$ & \text{} & \text{} & \text{} & \text{} 
 \\ \hline
 $\P_{\Delta H}$ & \text{} & $c_{\Delta H}$ & $a_{\Delta H}$ & \text{} & \text{} & \text{} & \text{} 
 \\ \hline
 $\P_{DH}$ & \text{} & \text{} & $c_{DH}$ & \text{} & \text{} & \text{} & \text{} 
 \\  \hline
\end{tabular}
\par\end{centering}
\caption{Illustration of which operators in $\L_4$ (see Eq.~(\ref{L4}) and Table~\ref{table:L4-terms}) contribute to \mbox{2-,} \mbox{3-,} and \mbox{4-point} amplitudes involving pions 
and/or $h$ fields. The specific  operator coefficients contributing to each amplitude are 
indicated, following the $c_i\cF_i$ expansion in Eq.~(\ref{Fi}). }
\label{table:ct-process}
\end{table}

\subsection{1-point functions}
\label{sec:1pt}
\begin{figure}
  \begin{center}
    \includegraphics[scale=0.35]{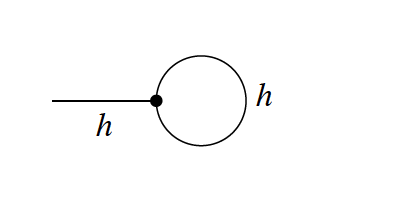}  
  \end{center}
  \caption{Diagram contributing to the Higgs 1-point function.}
  \label{fig:proc-h-tadpole}
\end{figure}

Because of chiral symmetry pions always come in even numbers in any vertex, unlike Higgs particles, thus tadpole contributions may be generated only for the latter. 
At tree-level it would suffice to set $\mu_1=0$ in $V(h)$ (Eq.~(\ref{L0})) in order to insure $\langle h \rangle=0$. At one-loop, a tadpole term is induced from the triple Higgs couplings $\mu_3$ and $a_H$, though, via 
 the Feynman diagram in Fig.~\ref{fig:proc-h-tadpole}.  The counterterm required to cancel this contribution reads 
\begin{equation}
\delta\mu_1^3=  m_h^2\paren{\mu_3-4a_H\frac{m_h^2}{v}} \div\,,
\end{equation}
and has no impact on 
the rest of the Lagrangian.

\subsection{2-point functions}
\label{sec:2pt}
\begin{figure}
  \begin{center}
    \includegraphics[height=5cm,width=\textwidth,keepaspectratio]{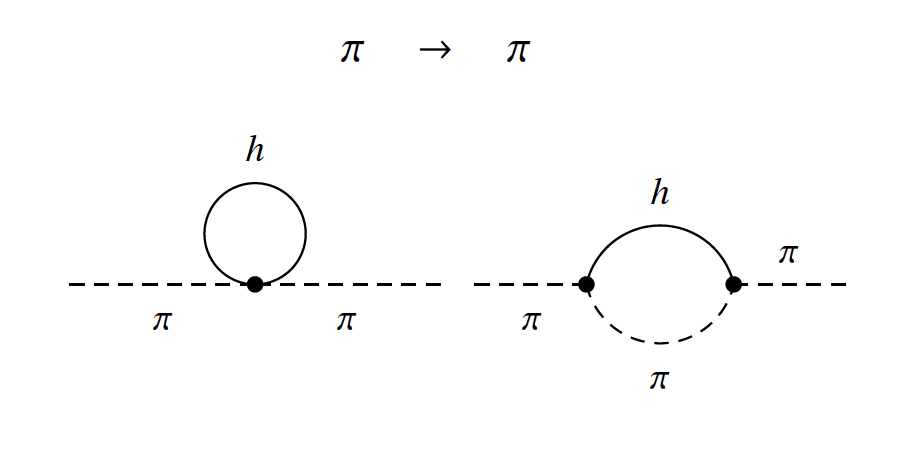}  
  \end{center}
  \caption{Diagrams contributing to the $\pi$ self-energy.}
  \label{fig:proc-pp}
\end{figure}

Consider mass and wave function renormalization for the pion and $h$
fields. Because of chiral symmetry no pion mass will be induced by
loop corrections at any order, unlike for the $h$ field, whose mass is
not protected by that symmetry.  The diagrams contributing to the pion
self-energy are shown in Fig.~\ref{fig:proc-pp}.
The divergent part of the amplitudes, $\Pi^{ij}_{\rm div}(p^2)\div$, and the
counterterm structure are given by
\begin{align}
\label{eq:div-pp}
 \Pi^{ij}_{\rm div}(p^2) =& \left[  p^2  \left(a_C^2-b_C\right) \frac{m_h^2}{v^2} + p^4 \frac{a_C^2}{v^2} \right]\delta_{ij}\,,
\\[3mm]
 \Pi^{ij}_{ctr}(p^2) =&  \left[
  p^2  \delta_\pi - p^4 \frac{4}{v^2}  \l\d c_{9} - \frac{\d v^2}{v^2}\r \right] \delta_{ij}\,.
\end{align}
In an off-shell renormalization scheme, it is necessary to match all the
momenta structure of the divergent amplitude with that of the counterterms,
which leads to the following determination
\begin{equation}
\begin{aligned}
\delta_\pi=&-\left(a_C^2-b_C\right) \frac{m_h^2}{v^2}\div\,,\\
\d c_{9} - \frac{\d v^2}{v^2} =& \frac{a_C^2}{4}\div\,.
\end{aligned}
\end{equation}
It follows that the $\pi$ wave function  renormalization has no divergent part whenever  
$a_C^2=b_C$, which happens for instance in the case of the SM ($a_C=b_C=1$).  
Note as well that the absence of a
constant term in eq.~(\ref{eq:div-pp}) translates into massless pions
at 1-loop level, as mandated by chiral symmetry 
at any loop order.  
Furthermore, the $p^4$ term stems from the
$h-\pi$ coupling $a_C$, which is an entire new feature compared to the 
 nonlinear $\sigma$ model renormalization.
  This term demands the presence of a
$\Box\bpi\Box\bpi$ counterterm in the $\L_4$ Lagrangian, as expected by naive
dimensional analysis.

\begin{figure}
  \begin{center}
    \includegraphics[height=5cm,width=\textwidth,keepaspectratio]{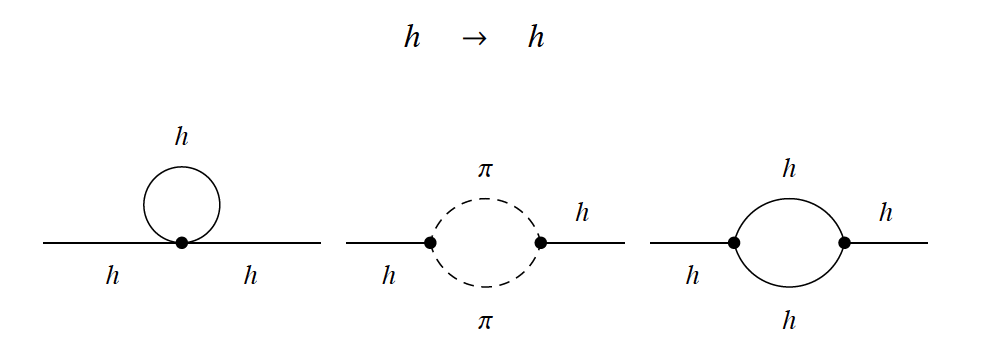}  
  \end{center}
  \caption{Diagrams contributing to the Higgs self-energy.}
  \label{fig:proc-hh}
\end{figure}

Turning to the Higgs particle, the diagrams contributing to its self-energy are
shown in Fig.~\ref{fig:proc-hh}, with the divergent part and the required counterterm structure given by
\begin{align}
\Pi_{\rm div}(p^2)=&
 p^4 \frac{ \left(3 a_C^2+a_H^2\right)}{2 v^2} +
p^2 \left(-\frac{\mu _3}{v} a_H+ \frac{m_h^2 \left(5 a_H^2-b_H\right)}{v^2}\right)
\nonumber\\+&\l
\frac{1}{2} \mu _3^2 +
  \frac{1}{2} m_h^2 \left(\lambda-8
   \frac{\mu _3}{v} a_H\right)
   +\frac{m_h^4 \left(6
   a_H^2-b_H\right)}{v^2}
    \r,
\\[3mm]
\Pi_{ctr}(p^2) =& p^4 \frac{2 \d c_{\Box H}}{v^2} + p^2 \d_h - \d m_h^2\,.
\end{align}
It follows that the required counterterms are given by 
\begin{equation}
\begin{aligned}
\delta _h =& \left[ \frac{\mu _3}{v} a_H+ \frac{m_h^2 \left(b_H-5
    a_H^2\right)}{v^2}\right] \div,
   \\
  \d m_h^2 =&
\left[ \frac{1}{2} \mu_3^2  +
\frac{1}{2} m_h^2 \left(\lambda-8 \frac{\mu _3}{v} a_H\right)+
   \frac{m_h^4 \left(6
   a_H^2-b_H\right)}{v^2}\right] \div,\\
\d c_{\Box H} =& - \frac{1}{4} \left(3 a_C^2+a_H^2\right)\div.
\end{aligned}
\end{equation}
This result implies that a non-vanishing $a_C$ (as in the SM limit) and/or $a_H$ leads to a
$p^4$ term in the counterterm Lagrangian, requiring a $\Box h\Box h$
term in $\L_4$. In this scheme, a Higgs wave function 
renormalization is operative only in deviations from the SM with non-vanishing $a_H$ and/or $b_H$.
\subsection{3-point functions}
\label{sec:3pt}
The calculational details for the 3- and 4-point functions will not be explicitly shown as they are not particularly
illuminating~\footnote{See Appendix~\ref{sec:AppA} for details and 
Ref.~\cite{SarasThesis} for an exhaustive description.}.  
Vertices with an odd number of legs necessarily involve at least one Higgs particle. 

\subsubsection*{$h h h$}
\begin{figure}
  \begin{center}
    \includegraphics[height=5.5cm,width=\textwidth,keepaspectratio]{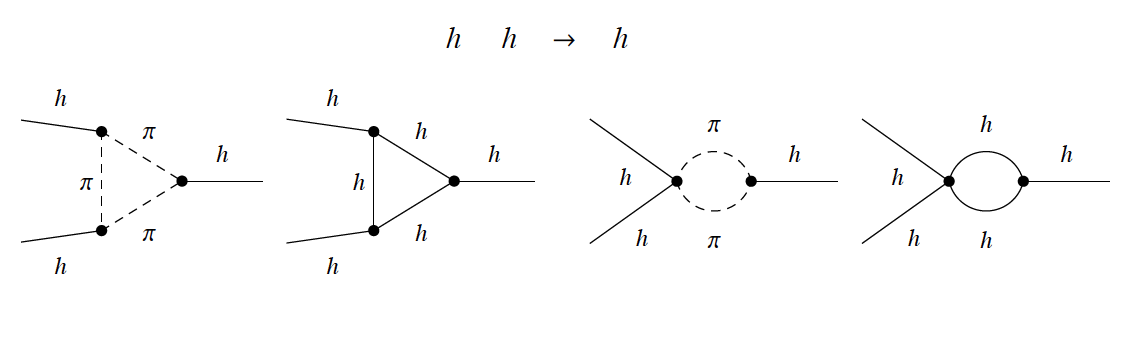}  
  \end{center}
  \caption{Diagrams contributing to the $hh\to h$ 
    amplitude, not including diagrams obtained by crossing.}
  \label{fig:proc-hhh}
\end{figure}

\vspace{-0.2cm}
\noindent
Let us consider first  the $hhh$ amplitude at one loop.  The
relevant diagrams to be computed are displayed in
Fig.~\ref{fig:proc-hhh}.
As $h$ behaves as a generic singlet, the vertices involving uniquely
external $h$ legs which appear in the Lagrangian Eq.~(\ref{L}) will
span all possible momentum structures that can result from one-loop
amplitudes. Hence any divergence emerging on amplitudes involving only
external $h$ particles will be easily absorbable.  The specific
results for the counterterms emerging from $\L_0$ and $\L_2$ can be
found in Appendix~\ref{sec:AppA}.

\subsubsection*{$\pi\pi h$}
\vspace{-0.2cm}
\noindent
The diagrams for $\pi\pi h$ amplitudes are shown in
Fig.~\ref{fig:proc-pph}.  The one-loop divergences are studied in
detail in Appendix~\ref{sec:AppA}; for instance, it turns out that
neither $\delta a_C$ nor $\delta a_{9}$ are induced in the SM limit.
Chiral symmetry restricts the possible structures spanned by the pure
$\pi$ and $h-\pi$ operators.  Because of this, it turns out that part
of the divergent amplitude induced by the last diagram in
Fig.~\ref{fig:proc-pph} cannot be cast as a function of the $\L_2$ and
$\L_4$ operators, that is, it cannot be reabsorbed by chiral-invariant
counterterms, and furthermore its coefficient depends on the pion
parametrization used: an apparent non chiral-invariant divergence
(NID) has been identified. NIDs are an artefact of the apparent
breaking of chiral symmetry when the one-loop analysis is treated in
perturbation theory~\cite{Weinberg:1968de} and have no physical impact
as they vanish for on-shell amplitudes. While long ago NIDs had been
isolated in perturbative analysis of four-pion vertices in the
non-linear sigma model~\cite{Appelquist:1980ae}, the result obtained
here is a new type of NID: a three-point function involving the Higgs
particle, corresponding to the chiral non-invariant operator
\begin{equation}
  \O_1^{\rm NID} = -a_C \l\frac{3}{2}+5\eta\r  \frac{\div}{v^3}\,\, \bpi\Box\bpi\Box h\,.\label{eq:NID-pph}\,
\end{equation}
This coupling cannot be reabsorbed as part of a chiral invariant
counterterm, but its contribution to on-shell amplitudes indeed vanishes.
  It is interesting to note that while the renormalization
conditions of all physical parameters turn out to be independent of
the choice of $\U$ parametrization, as they should, NIDs exhibit
instead an explicit $\eta$ dependence, as illustrated by 
Eq.~(\ref{eq:NID-pph}).  This pattern will be also present in the
renormalization of 4-point functions, developed  next.
\begin{figure}
  \begin{center}
    \includegraphics[height=7cm,width=\textwidth,keepaspectratio]{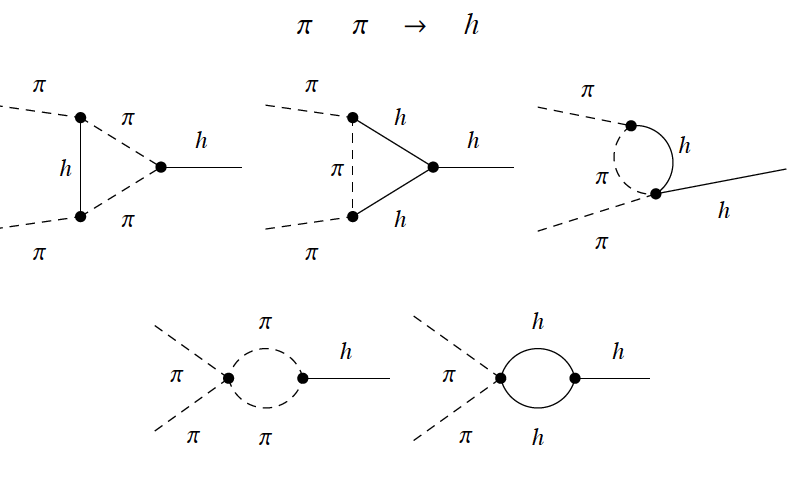}  
  \end{center}
  \caption{Diagrams contributing to the $\pi\pi\to h$ scattering
    amplitude, not including diagrams obtained by crossing.}
  \label{fig:proc-pph}
\end{figure}

\subsection{4-point functions}
\label{sec:4pt}
The analysis of this set of correlation functions turns out to be
tantalizing when comparing the results for mixed $\pi-h$ vertices with
those for pure pionic ones~\footnote{ It provides in addition 
  nice checks of the computations; for instance we checked explicitly
  in the present context that the consistency of the renormalization
  results for four-point functions requires $\delta_v^2=0$.}.

\subsubsection*{$\pi\pi h h$}
\vspace{-0.2cm}
\noindent
The computation of the $\pi\pi\to h h$ one-loop amplitude shows that the renormalization procedure requires the presence of  all possible chiral invariant 
$hh\pi\pi$ counterterms in the Lagrangian, in the most general
case.

Furthermore, we have  identified new NIDs in   $hh\pi\pi$
amplitudes: 
\begin{equation}\label{eq:NID-pphh}
  \begin{aligned}
  &\O^{\rm NID}_{2} = + (2a_C^2-b_C) \l \frac{3}{2}+5\eta\r \frac{\div}{v^4}\, \,
    \bpi\Box\bpi\, h\Box h\,,\\
  &\O^{\rm NID}_{3} = + (a_C^2-b_C)\l \frac{3}{2}+5\eta\r \frac{\div}{v^4}\, \,
    \bpi\Box\bpi \,\der_\mu h\der^\mu h\,,\\
  &\O^{\rm NID}_{4} = - 2a_C^2 \l \frac{3}{2}+5\eta\r\frac{\div}{v^4}\,\,
    \bpi\der_\mu\bpi\,\der^\mu h\Box h\,.
  \end{aligned}
\end{equation}
While these NIDs differ from that for the three-point function in
Eq.~(\ref{eq:NID-pph}) in their counterterm structure, they all share
an intriguing fact: to be proportional to the factor
$(3/2+5\eta)$. Therefore a proper choice of parametrization,
i.e. $\eta=-3/10$, removes all mixed $h-\pi$ NIDs.  That value of
$\eta$ is of no special significance as fas as we know, and in fact
there is no choice of parametrization that can avoid {\it all}
noninvariant divergencies, as proved next.
\noindent

\subsubsection*{$\pi\ppi\pi\pi$} 
\vspace{-0.2cm}
\noindent
Consider now $\pi\pi\to\pi\pi$ amplitudes.  Only two counterterms are
necessary to reabsorb chiral-invariant divergencies, namely $\d c_{6}$
and $\d c_{11}$.  In this case, we find no other NIDs than those
already present in the nonlinear $\sigma$
model~\cite{Appelquist:1980ae}, which stemmed from the insertion in
the loop of the four-pion vertex (whose coupling depends on $\eta$).
Our analysis shows that the four-$\pi$ NIDs read:
\begin{equation}\label{eq:NID-pppp}
  \begin{aligned}
  &\O_5^{\rm NID} = +\left(9 \eta^2 + 5 \eta + \frac{3}{4}\right)\frac{\div}{v^4}\,
                  (\bpi\Box\bpi)^2,  \\
  &\O_6^{\rm NID} = +\left[ 1+4\eta+\l \frac{1}{2}+\eta \r a_C^2 \right] 
                  \frac{\div}{v^4}\,(\bpi\Box\bpi)(\der_\mu\bpi\der^\mu\bpi),\\
  &\O_7^{\rm NID} = +2\eta^2 \frac{\div}{v^4}\,\bpi^2(\Box\bpi)^2,\\
  &\O_8^{\rm NID} = +2\eta \left(a_C^2-1\right)\frac{\div}{v^4}\,
                 (\Box\bpi\der_\mu\bpi)(\bpi\der^\mu\bpi).
  \end{aligned}
\end{equation}
As expected, the parametrization freedom -- the dependence on the
$\eta$ parameter -- appears only in NIDs, and never on
chiral-invariant counterterms, as the latter describe physical
processes.  Furthermore, the contribution of all NIDs to on-shell
amplitudes vanishes as expected~\footnote{This is not always seen when
  taken individually. For instance, the contribution of $\O_4^{\rm
    NID}$ to the $hh\pi\pi$ amplitude is cancelled by that of
  $\O_1^{\rm NID}$, which corrects the $h\pi\pi$ vertex.}.  Finally,
the consideration of the ensemble of three and four-point NIDs in
Eqs.~(\ref{eq:NID-pph}), (\ref{eq:NID-pphh}) and (\ref{eq:NID-pppp})
shows inmediately that no parametrization can remove {\it all} NIDs:
it is possible to eliminate those involving $h$~\footnote{This may be
  linked to the larger freedom of redefinition for fields not subject
  to chiral invariance.}, but no value of $\eta$ would remove all pure
pionic ones.

\subsubsection*{$hhhh$}
\vspace{-0.2cm}
\noindent
The
renormalization procedure  for $h h\to h h$ amplitudes is straightforward. It results in contributions to 
$\d a_{\Delta H}$, $\d c_{DH}$, and $\d b_{H}$. Interestingly,  Appendix~\ref{sec:AppB} 
illustrates that  large coefficients are present in some  terms of the RGE for $b_H$ and
$\lambda$; this might {\it a priori} translate into measurable effects when comparing data at different scales, if ever deviations from the SM predictions are detected, see Sect.~(\ref{sec:RGEs}).  

\begin{center}
\line(1,0){50}
\end{center}

There is a particularity of the off-shell renormalization scheme which deserves to be pointed out. A closer look at the
counterterms reveals that, in the SM case, that is
\begin{equation}
a_C=b_C=1, \quad a_H=b_H=0, \quad \mu_3=3 \frac{m_h^2}{v} \quad and
\quad \lambda = 3 \frac{m_h^2}{v^2},
\end{equation}
several BSM operator coefficients do not vanish. Although at first
this might look counterintuitive,  when
calculating physical amplitudes the contribution of these
non-vanishing operator coefficients all combine in such a way that the overall BSM
contribution indeed cancels. The same pattern propagates to the renormalization group equations discussed in Sect.~4.

\subsection{Dealing with the apparent non-invariant divergencies}
\label{sec:NID-counterterms}
 For the nonlinear $\sigma$ model the issue of NIDs was
analyzed long ago~\cite{Charap:1970xj, Kazakov:1976tj, Kazakov:1977mw,
  deWit:1979gw, Honerkamp:1971sh, Appelquist:1980ae}).  In that case,
it was finally proven that a nonlinear redefinition of the pion field which
includes space-time derivatives could reabsorb
them~\cite{Appelquist:1980ae}.  This method reveals a deeper rationale
in understanding the issue, as Lagrangians related by a local
field redefinition are equivalent, even when it involves
derivatives~\cite{ostrogradsky1850memoire, GrosseKnetter:1993td,
  Scherer:1994wi, Arzt:1993gz}. Consequently, if via a general pion
field redefinition
\begin{equation*}
  \bpi\to \bpi\, f(\bpi,h,\dmu\bpi,\dmu h,\dots)\,,
\end{equation*}
with $f(0)=1$, the Lagrangian is shifted 
\begin{equation*}
  \L\to \L^{\prime}=\L+\d \L\,,
\end{equation*}
from the equivalence between $\L$ and $\L^\prime$ it follows that 
$\d\L$  must be unphysical. Thus, if  an
approppriate pion field redefinition is found which is able to absorb all NDIs, it automatically implies that NDIs do not contribute to the
$S$-matrix, and therefore that chiral symmetry remains
unbroken.  {\it In other words, the non-invariant
  operators can be identifed with quantities in the functional generator that vanish
  upon performing the path integral.}

Let us consider the following pion redefinition, in which we propose new terms not considered previously and which contain the $h$ field: 
\begin{equation} \label{pionredefgen}
\begin{aligned}
\pi_i \to&\, \pi_i \l 1+ \frac{\a_1}{2v^4}\bpi\Box\bpi 
   + \frac{\a_2}{2v^4} \der_\mu\bpi\der^\mu\bpi+ \frac{\b}{2v^3}\Box h 
   + \frac{\tilde{\g}_1}{2v^4}h\Box h+ \frac{\g_2}{2v^4}\der_\mu h\der^\mu h \r\\
   &\qquad + \frac{\alpha_3}{2v^4} \Box \pi_i (\bpi\bpi) 
   + \frac{\alpha_4}{2v^4}\der_\mu\pi_i (\bpi\der^\mu\bpi).    
\end{aligned}
\end{equation} 
The application of this redefinition to $\L_4$ is immaterial, as it would only induce couplings of higher order.  As all terms in the shift contain two derivatives, when  applied to  $\L_2$ contributions to  $\L_4$  and  NID operator coefficients do follow.  Indeed, the action of Eq.~(\ref{pionredefgen}) on $\L_2$ reduces to that on the term
\begin{equation}
  \frac{1}{4}\Tr(\der_\mu \U \der^\mu\U)\F_C(h)\,,
\end{equation}
which produces the additional contribution to NDI vertices given by
\begin{align}
&\Delta\L^{\rm NID} = - \bpi\Box\bpi \l 
\frac{\a_1}{v^4}\bpi\Box\bpi + \frac{\a_2}{v^4} \der_\mu\bpi\der^\mu\bpi  +
 \frac{\b }{v^3} \Box h  +  \frac{\g_1}{v^4}h\Box h
+  \frac{\g_2}{v^4}\der_\mu h\der^\mu h   \r \nn
 \\  
&\qquad\quad\!  - \frac{\alpha_3}{v^4} (\Box\bpi\Box\bpi)(\bpi\bpi) - \frac{\alpha_4}{v^4} (\Box\bpi\der_\mu\bpi)(\bpi\der^\mu\bpi)
      - \frac{2 a_C \b}{v^4} \bpi\der_\mu\bpi\der^\mu h \Box h + ... 
\end{align}
where $\gamma_1 = 2a_C\beta+\tilde\gamma_1$, and where the dots indicate other operators with either six derivatives
 or that have more than four fields and are beyond the scope of this paper. 
 Comparing the terms in $\Delta\L^{\rm NID}$ with the NID operators
found,  Eqs.~(\ref{eq:NID-pph}),
(\ref{eq:NID-pphh}) and (\ref{eq:NID-pppp}), it follows that by choosing 
\begin{equation*}
  \begin{aligned}
    &\a_1 = \left(9 \eta^2 + 5 \eta + \frac{3}{4}\right)\div, \\
    &\a_2 = \left[ 1+4\eta+\l \frac{1}{2}+\eta \r a_C^2 \right]  \div,\\
    &\a_3 = 2\eta^2 \div,\\
    &\a_4 = 2\eta \left(a_C^2-1\right)\div,
  \end{aligned}\qquad
  \begin{aligned}
    &\beta = -\l\frac{3}{2}+5\eta\r a_C\div,\\
    &\gamma_1 = \l\frac{3}{2}+5\eta\r \left(2 a_C^2-b_C\right)\div, \\
    &\gamma_2 = \l\frac{3}{2}+5\eta\r \left(a_C^2-b_C\right)\div.\\
  \end{aligned}
\end{equation*}
all 1-loop NID are removed away.

A few comments are in order.
The off-shell renormalization of one loop
amplitudes  is delicate and physically interesting.  Because of chiral
symmetry, the pure pionic or mixed pion-$h$  operators do not encode all
possible momentum structures, even after pion field redefinitions.
Hence, the appearance of divergent structures that can be absorbed by
$\d\L_0$, $\d\L_2$, $\d\L_4$ and $\Delta\L^{\rm NI}$ is a
manisfestation of chiral invariance and  of the field redefinition
equivalence discussed above.   
We have shown consistently that  NIDs
 appearing in the one-loop renormalization of
the electroweak chiral Lagrangian do not contribute to on-shell
quantities.  In fact, a closer examination has revealed that the
apparent chiral non-invariant divergencies emerge from loop diagrams which have at least
one four-pion vertex in it, and this is why all of them depend on
$\eta$. We have also shown that the presence of a light Higgs boson modifies the coefficients of the unphysical counterterms  made out purely of pions, 
but not their structure,  
neither -of course- breaks chiral symmetry.

\section{ Renormalization Group Equations} 
\label{sec:RGEs}
It is straightforward to derive the RGE from the $\delta c_i$ divergent contributions determined in the previous section. The complete  RGE set can be found in Appendix~\ref{sec:AppB}. 
As illustration,  the evolution of those Lagrangian coefficients which do not vanish in the  SM limit is given by:
\begin{align}
  \mudmu a_C&= \frac{1}{2}a_C\left[a_H \frac{\mu_3}{v} + \left(3 b_C-5 a_H^2+b_H\right)\frac{m_h^2}{v^2}\right] 
               +a_C^2\left(\frac{\mu _3}{2v}-2 a_H \frac{m_h^2}{v^2}\right) 
               -\frac{3}{2}a_C^3\frac{m_h^2}{v^2}\nn\\
               &-\frac{1}{2v}b_C \mu_3+2 a_H b_C \frac{m_h^2}{v^2} \,,\\
 \mudmu  b_C&= b_C \left[2 a_C \frac{\mu_3}{v}+5 a_H \frac{\mu_3}{v}-\frac{\lambda}{2}
               -\left(5 a_C^2+8 a_H a_C+17 a_H^2-3 b_H\right) \frac{m_h^2}{v^2}\right]
             +b_C^2\frac{ m_h^2}{v^2}\nn\\
            &+\frac{1}{2}\left(-4 a_C \frac{\mu_3}{v}-8 a_H \frac{\mu_3}{v}+\lambda \right) a_C^2
            +2 \left(2 a_C^2+4 a_H a_C+6 a_H^2-b_H\right)a_C^2 \frac{m_h^2 }{v^2}\nn\\
             \mudmu m_h^2&=-\frac{1}{2}  \mu_3^2 
              +\left(5 a_H \frac{\mu_3}{v}-\frac{\lambda}{2}\right) m_h^2
              + \left(2 b_H-11 a_H^2\right) \frac{m_h^4}{v^2} \,,\\
 \mudmu \mu_3&=\frac{1}{2}\mu_3\left[\left(-a_C^2+b_C-87 a_H^2+15 b_H\right) 
                  \frac{m_h^2}{v^2}-3 \lambda\right]+\frac{15}{2v} \mu_3^2a_H\nn\\
                 & +6 a_H \lambda \frac{m_h^2}{v}
                 +6 \left(8 a_H^3-3 a_H b_H\right) \frac{m_h^4}{v^3}\,,\\
 \mudmu \lambda&=\lambda\left[26 a_H \frac{\mu_3}{v}
                                  +\left(14 b_H-82 a_H^2\right) \frac{m_h^2}{v^2}\right]
                  -\frac{3}{2}\lambda^2 +12\left(b_H-6 a_H^2\right) \frac{\mu_3^2}{v^2}\\
                  &+48 a_H \left(8 a_H^2-3 b_H\right) \mu_3 \frac{m_h^2}{v^3}
                  -6 \left(80 a_H^4-48 b_H a_H^2+3 b_H^2\right) \frac{m_h^4}{v^4}\,. \label{lambdaRGE}
                  \end{align}
These and the rest of the RGE in Appendix~\ref{sec:AppB} show as well
that the running of the parameters $a_C$, $b_C$, $a_H$, $b_H$, and
$v^2$ is only induced by the couplings entering the Higgs potential,
Eq.~(\ref{V}).

Note that in the RGE for the Higgs quartic self-coupling $\lambda$,
Eq.~(\ref{lambdaRGE}), some terms are weighted by numerical factors of
${\cal O} (100)$. This suggests that if a BSM theory
results in small couplings for $a_H$ and $b_H$, those
terms could still induce measurable phenomenological consequences. Nevertheless, physical
amplitudes will depend on a large combination of parameters, which
might yield cancellations or enhancements as pointed out earlier, and only a more thorough
study can lead to firm conclusions. Such  large
coefficients turn out to be also present in the evolution of some BSM couplings,
such as the four-Higgs coupling $b_H$ for which
\begin{align} \label{b_H}
 \mudmu b_H&=b_H \left[20 a_H \frac{\mu_3}{v}-\frac{3}{2}\lambda
                   + \left(-a_C^2+b_C-87a_H^2\right) \frac{m_h^2}{v^2}\right]\nn\\
                   &-42 \frac{\mu_3}{v} a_H^3
                   +\frac{13}{2} \lambda a_H^2
                   +\left(7 b_H^2+120 a_H^4\right)\frac{m_h^2}{v^2} \,.    
\end{align}
On general grounds $a_H$ is expected to be small, and for instance  the $a_H^4$ 
dependence in Eq.~(\ref{b_H}) is not expected to be relevant in spite
of the numerical prefactor. On the other side, present data set
 basically no bound on the couplings involving three or more
external Higgs particles, and thus the future putative impact of this evolution
should not be dismissed yet.

\section{Comparison with the literature}
\label{sec:comparison}

Previous works on the one-loop renormalization of the scalar sector of the non-linear Lagrangian with a light Higgs have used either the square root parametrization ($\eta=0$ in our parametrization) or the exponential one ($\eta=-1/6$), producing very interesting results, and have
\begin{itemize}
\item concentrated on on-shell analyses,
\item disregarded the impact of $\mathcal{F}_H(h)$,
 \item disregarded fermionic operators; in practice this means to neglect all fermion masses.
 \end{itemize}
This last point is not uncorrelated with the fact that the basis of
independent four-derivative operators determined here has a larger
number of elements than previous works about  the scalar sector. Those extra bosonic operators have
been shown here to be required by the counterterm procedure.  It is
possible to demonstrate, though, that they can be traded via EOM by
other type of operators including gauge corrections and Yukawa-like
operators. In a complete basis of all possible operators it is up to
the practitioner to decide which set is kept, as long as it is
complete and independent. When restricting instead to a given subsector,
the complete and consistent treatment requires to consider all independent operators
of the kind selected (anyway the renormalization procedure will
indicate their need), or to state explicitly any extra assumptions to
eliminate them. Some further specific comments:

Ref.~\cite{Delgado:2013hxa} considers, under the first two itemized
conditions above plus disregarding the impact of $V(h)$ (and in
particular neglecting the Higgs mass), the scattering processes $hh\to
hh$, $\pi\pi \to hh$ and $\pi\pi \to \pi\pi$. With the off-shell
treatment, five additional operators result in this case with respect
to those obtained in that reference (assuming the rest of their
assumptions), $\mathcal{P}_7$, $\mathcal{P}_9$, $\mathcal{P}_{10}$,
$\mathcal{P}_{\Box H}$ and $\mathcal{P}_{\Delta H}$ in
Table~\ref{table:L4-terms}.  Note that all these operators contain
either $\Box h$ or $\Box\pi$ inside; they may thus be implicitly
traded by fermionic operators via EOM, and can only be disregarded if
all fermion masses are neglected. Assuming this extra condition, we
could reproduce their results using the EOMs.  For instance, the RGEs
derived here for $c_{6}$, $c_{8}$, $c_{20}$, $a_C$ and $b_C$ differ
from the corresponding ones in that reference: an off-shell
renormalization analysis entails the larger number of operators
mentioned.  In any case, we stress again that the results of both
approaches coincide when calculating physical amplitudes. Another
contrast appears in the running of $a_C$, $b_C$, $a_H$, $b_H$, as well
as the mass, the triple, and the quartic coupling of the Higgs, for
which the running is induced by the Higgs potential parameters,
disregarded in that reference.

In Ref.~\cite{Espriu:2013fia} the on-shell scattering process of the
longitudinal modes of the process $W^+W^- \rightarrow ZZ $ is
considered, disregarding $\mathcal{F}_H(h)$ but including the impact
of $V(h)$.  Our off-shell treatment results in this case in one
additional operator (assuming the rest of their assumptions) with
respect to those in that reference, $\mathcal{P}_9$, as only processes
involving four goldstone bosons were considered there. Again this
extra operator contains $\Box\pi$ in all its terms and it could be
neglected in practice if disregarding all fermion masses. With this
extra assumption, our results reduce to theirs in the limit
indicated.

\section{Conclusions}
\label{sec:conclusions}

We have considered the one-loop off-shell renormalization of  the effective non-linear Lagrangian in the presence of a light (Higgs) scalar particle, up to four-derivative terms and taking into account the finite Higgs mass. We have concentrated in its scalar sector: goldstone bosons (that is, the longitudinal components of the SM gauge bosons) and the light scalar $h$.

Analyzing the custodial-preserving sector, we have determined the 
four-derivative counterterms required by the one-loop renormalization
procedure, by considering the full set of 1-, 2-, 3- and 4-point
functions involving pions and/or $h$.  The off-shell treatment has
allowed to determine all required counterterms, confirming for the
sector analyzed that the generic low-energy effective non-linear
Lagrangian with a light Higgs particle developed in
Refs.~\cite{Alonso:2012px, Brivio:2013pma} is complete: all
four-derivative operators of that basis and nothing else is induced by
the renormalization. Those operators are linearly independent and form
thus a complete basis when that sector is taken by itself.  It is
shown that a larger number of operators than previously considered are then needed.

As a useful analysis tool, we have also proposed here a general parametrization for the goldstone boson matrix, which at the order considered here depends on only one parameter $\eta$,  and reduces to the popular parametrizations (square root, exponential etc.) for different values of $\eta$. 
All counterterms induced by the renormalization procedure are then easily seen to be parametrization independent, as it befits physical couplings. 

Furthermore, new chiral non-invariant counterterms involving the Higgs
particle and pions have been found in our perturbative analysis. These
findings extend to the realm of the Higgs particle the apparent non
chiral-invariant divergences identified decades ago for the non-linear
sigma model~\cite{Appelquist:1980ae}. Those apparent violations of
chiral symmetry are an artefact of perturbative approaches, they
vanish on-shell, and their origin had been tracked down to the
insertion of the four-pion vertex in loops.  In this paper, new
non-invariant divergences are shown to appear in triple $h\pi\pi$
counterterms and in $hh\pi\pi$ ones, and shown to have the same
origin.  Interestingly: i) all apparently non-invariant divergences
depend explicitly on $\eta$, consistent with their non-physical
nature; ii) there is a value of the $\eta$ parameter for which the
non-invariant divergences involving the Higgs vanish, though, while no
$\eta$ value can cancel the ensemble of non-invariant divergences and
in particular the pure pion ones.

 Moreover, we have determined a local pion-field redefinition which
 includes space-time derivatives and reabsorbs automatically {\it all}
 apparently chiral non-invariant counterterms. This field redefinition
 leaves invariant the S-matrix and thus the result shows automatically
 that chiral symmetry remains unbroken.

 For the physical counterterms induced, we observe a complete
 agreement with the naive dimensional
 analysis~\cite{Manohar:1983md,Jenkins:2013sda} in the $h-\pi$ sector
 of the chiral Lagrangian. Finally, the RGEs for the scalar sector of
 the general non-linear effective Lagrangian for a Higgs particle have
 been also derived in this work. The complete set of equations can be
 found in Appendix~\ref{sec:AppB}.  Factors of ${\cal O}(100)$ appear
 accompanying certain operator coefficients in the RGEs, and those
 terms may thus be specially relevant when comparing future Higgs and
 gauge boson data obtained at different energies. On more general
 grounds, although present data are completely consistent with the SM
 predictions, going for precision in constraining small parameters may
 be the best way to tackle BSM physics and we should not be deterred
 by the task: the dream of today may be the discovery of tomorrow and
 the background of the future.

\section*{Acknowledgements}
We acknowledge illuminating conversations with Ilaria Brivio,
Ferruccio Feruglio, Howard Georgi, Mar\'ia Jos\'e Herrero, Luca Merlo,
Pilar Hern\'andez and Stefano Rigolin.  We also acknowledge partial
support of the European Union network FP7 ITN INVISIBLES (Marie Curie
Actions, PITN-GA-2011-289442), of MICINN, through the project
FPA2012-31880, and of the Spanish MINECO's ``Centro de Excelencia
Severo Ochoa'' Programme under grant SEV-2012-0249.  The work of
K.K. and P.M. is supported by an ESR contract of the European Union
network FP7 ITN INVISIBLES mentioned above. The work of S.S. is
supported through the grant BES-2013-066480 of the Spanish of
MICINN. K.K. acknowledges IFT-UAM/CSIC for hospitality during the
initial stages of this work.

\appendix
\section{The counterterms}
\label{sec:AppA}
Details about the computation of the counterterms and the
renormalization of the chiral Lagrangian are given in this Appendix, including the derivation of the RGEs. 

The bare parameters (denoted
by $b$) written in terms of the renormalized ones and the counterterms
for the $\L_2$ and $\L_4$ Lagrangians are given  by 
\begin{equation}
\begin{aligned}
&h_b = \sqrt{Z_h} h, \qquad \d_h \equiv Z_h-1, \\
&\bpi_b = \sqrt{Z_\pi}\bpi,   \qquad \d_\pi \equiv Z_\pi-1, \\ \\
&v^2_b = Z_\pi (v^2+\d v^2)\mu^{-\eps}, \\
&(m_h^2)^b = \frac{1}{Z_h}(m_h^2+\delta m_h^2),\\
&(\mu_1^3)^b = \frac{1}{Z_h^{1/2}} \l \mu_1^3 + \delta \mu_1^3 \r\mu^{3\eps/2},\\
&\mu_3^b = \frac{1}{Z_h^{3/2}} \l \mu_3 + \delta \mu_3 \r\mu^{\eps/2},\\
&\lambda^b = \frac{1}{Z_h^{2}} \l \lambda + 
                      \delta \lambda\r\mu^{\eps},
\end{aligned}
\quad\,\,
\begin{aligned}
\\ \\
&a_C^b = \frac{1}{Z_\pi^{1/2}Z_h^{1/2}} 
              \l a_C+\delta a_C+\frac{a_C}{2}\frac{\delta v^2}{v^2} \r, \\
&b_C^b = \frac{1}{Z_h} \l b_C+\delta b_C+b_C\frac{\delta v^2}{v^2} \r, \\
&a_H^b = \frac{Z_\pi^{1/2}}{Z_h^{3/2}} \l  a_H+\delta a_H +\frac{a_H}{2}\frac{\delta v^2}{v^2}\r, \\
&b_H^b = \frac{Z_\pi}{Z_h^{2}} \l b_H+\delta b_H +b_H\frac{\delta v^2}{v^2}\r,
\end{aligned}
\end{equation}
where 
\begin{equation}
\begin{aligned}
&X_i^b = \l X_i + \delta X_i + 2X_i \frac{\delta v^2}{v^2}\r\mu^{-\eps}, &X_i=c_{6},c_{9},c_{11},
  \\
&X_i^b = \frac{Z_\pi^{1/2}}{Z_h^{1/2}}\l X_i + \delta X_i 
                + \frac{3}{2}X_i \frac{\delta v^2}{v^2} \r\mu^{-\eps}, &X_i=c_{7},a_{9},c_{10},
                \\
&X_i^b = \frac{Z_\pi}{Z_h}\l X_i + \delta X_i + 2X_i \frac{\delta v^2}{v^2} \r\mu^{-\eps}, &X_i=a_{7},c_{8},b_{9},a_{10},c_{20},\\
&X_i^b = \frac{Z_\pi}{Z_h}\l X_i + \delta X_i + X_i \frac{\delta v^2}{v^2} \r\mu^{-\eps}, &X_i=c_{\Box H},\\
&X_i^b = \frac{Z_\pi^{3/2}}{Z_h^{3/2}}\l X_i + \delta X_i + \frac{3}{2}X_i \frac{\delta v^2}{v^2} \r\mu^{-\eps}, &X_i=a_{\Box H},c_{\Delta H},\\
&X_i^b = \frac{Z_\pi^{2}}{Z_h^{2}}\l X_i + \delta X_i + 2X_i \frac{\delta v^2}{v^2} \r\mu^{-\eps}, &X_i=b_{\Box H},a_{\Delta H},c_{DH}.
\end{aligned}
\end{equation}

\begin{figure}
  \begin{center}
    \includegraphics[width=\textwidth,keepaspectratio]{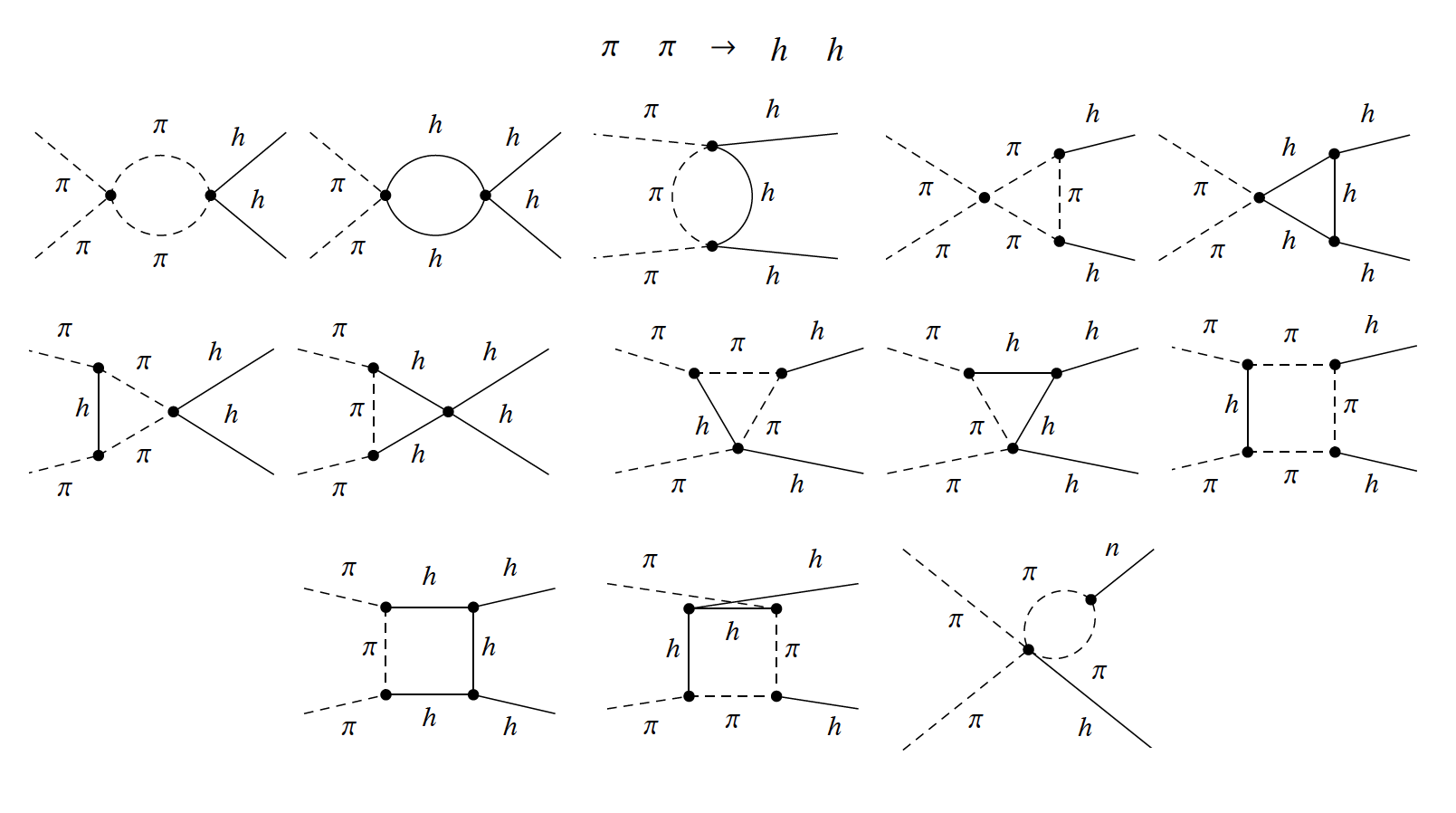}  
  \end{center}
  \caption{Diagrams contributing to the $\pi \pi \to hh$ 
    amplitude, not including diagrams obtained by crossing.}
  \label{fig:proc-pphh}
\end{figure}

\begin{figure}
  \begin{center}
    \includegraphics[scale=0.35]{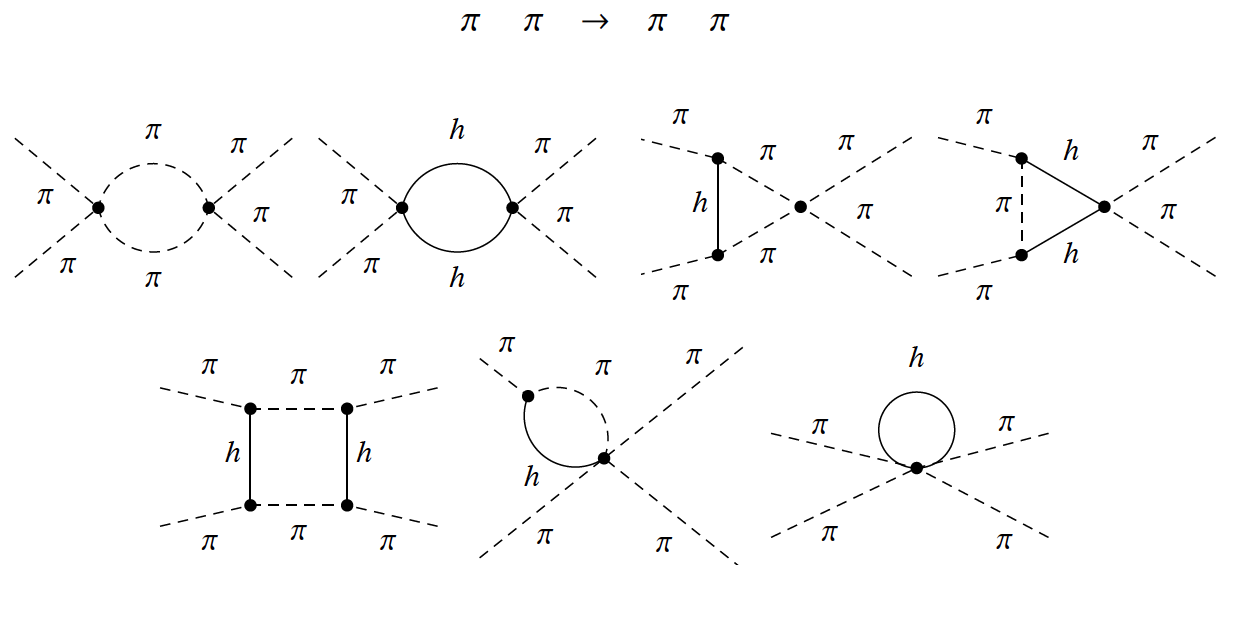}  
  \end{center}
  \caption{Diagrams contributing to the $\pi \pi \to \pi \pi$ 
    amplitude, not including diagrams obtained by crossing.}
  \label{fig:proc-pppp}
\end{figure}

\begin{figure}
  \begin{center}
    \includegraphics[height=8cm,width=\textwidth,keepaspectratio]{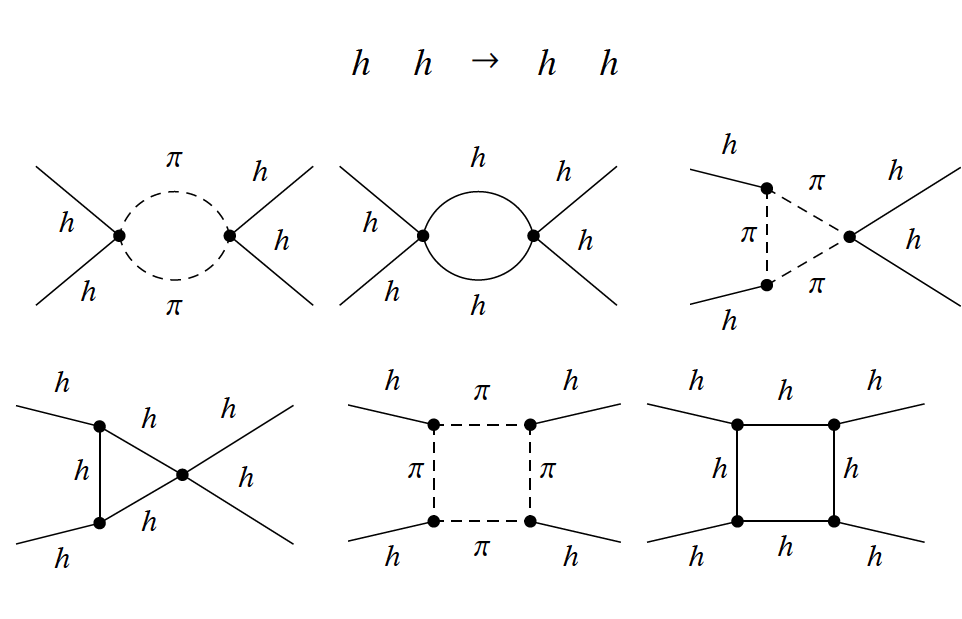}  
  \end{center}
  \caption{Diagrams contributing to the $hh\to hh$ 
    amplitude, not including diagrams obtained by crossing.}
  \label{fig:proc-hhhh}
\end{figure}

The counterterms required to absorb the divergencies of the $hhh$
3-point function are
\begin{equation}
  \begin{aligned}
    &\d a_{\Box H}= \frac{1}{2}\l-\frac{3 a_C b_C}{2}-\frac{a_H b_H}{2}+3 a_C^3+a_H^3\r\div, \\
    &\d c_{\Delta H}= \frac{1}{2} \left(-3 a_C b_C+3 a_C^3-a_H^3\right)\div, \\
    &\d a_{H}= \left[\frac{1}{2} \left(-9 \frac{\mu_3}{v}
      a_H^2+\lambda a_H+2 \frac{\mu_3}{v} b_H\right) +  a_H\left(15 a_H^2-7 b_H\right) 
		\frac{ m_h^2}{v^2}\right]\div, \\
    &\d \mu_3= \bigg[\frac{3}{2} \mu_3 \left(\lambda -4
      \frac{\mu_3}{v} a_H\right)+ 6 \left(6
      \mu_3 a_H^2-\lambda v a_H-\mu_3 b_H\right) \frac{ m_h^2 }{v^2} \\ 
      & \hspace{6.6cm} + 6 a_H  \left(3 b_H-8 a_H^2\right)  \frac{m_h^4}{v^3}\bigg]\div,
  \end{aligned}
\end{equation}
while those for $\pi\pi\to h$ read
\begin{equation}
\begin{aligned}
 &\d a_C = \frac{1}{2} \left(a_C^2-b_C\right) \left[2(a_C+2a_H)\frac{ m_h^2 }{v^2}-\frac{\mu_3}{v}\right]\div, \\
 &\d c_{7} = \frac{1}{4} \left(-a_H b_C+a_C^2 a_H-a_C^3-2 a_C\right)\div,\\
 &\d a_{9} = -\frac{1}{8} a_C \left(a_C a_H+a_C^2-b_C\right)\div, \\
 &\d c_{10} = \frac{1}{2} a_C \left(-a_C a_H+a_C^2+b_C\right)\div.
\end{aligned}
\end{equation}
In the case of the $\pi\pi\to h h$ amplitudes, 
 the
relevant diagrams are displayed in
Fig.~\ref{fig:proc-pphh}, and 
the counterterms correspond to
\begin{equation}
\begin{aligned}
  &\d b_C = \frac{1}{2} \left(a_C^2-b_C\right) 
	\Big[\left(4 a_C+8 a_H\right)\frac{\mu_3}{v}-\lambda  \\
    &\hspace{4cm}-2\left(8 a_C a_H+4 a_C^2+12 a_H^2-b_C-2 b_H\right) \frac{ m_h^2 }{v^2} 
    \Big]\div, \\
  &\d a_{7}= \frac{1}{8} \Big[a_C^2 \left(-4 a_H^2-3 b_C+b_H+4\right)+2 a_C a_H b_C
	+b_C \left(4 a_H^2-b_H-2\right)+4a_C^4\Big]\div\,, \\
  &\d c_{8} = \frac{1}{3} \left[a_C^2 \left(a_H^2+b_C\right)-2 a_C a_H b_C-a_C^3 a_H+a_C^4+b_C^2\right]\div\,, \\
  &\d b_{9} = \frac{1}{4} \left[-a_C^2 \left(-4 a_H^2+5 b_C+b_H\right)-4 a_C a_H b_C+4 a_C^3 a_H+4 a_C^4+b_C^2\right]\div\,, \\
  &\d a_{10}= \frac{1}{4} \left[a_C^2 \left(4 a_H^2+b_C-b_H\right)-4 a_C a_H b_C-4 a_C^4+b_C^2\right]\div\,, \\
&\d c_{20}=  \frac{1}{12} \Big[a_C^2 \left(2 a_H^2-b_C+6\right)+2 a_C a_H b_C\\
    &\hspace{5cm} -b_C \left(3 a_H^2+b_C+6\right) -2 a_C^3 a_H+2 a_C^4\Big]\div\,.
\end{aligned}
\end{equation}
For $\pi\pi\to\pi\pi$ amplitudes, the 
relevant diagrams are displayed in
Fig.~\ref{fig:proc-pppp}, and 
the required counterterms are given by
\begin{equation}
\begin{aligned}
  &\d c_6 = \frac{1}{48} \left[a_C^2 \left(6 b_C-8\right)-2 a_C^4-3 b_C^2-2\right]\div, \\
  &\d c_{11}= -\frac{1}{12} \left(a_C^2-1\right)^2\div.
\end{aligned}
\end{equation}
Finally, the relevant diagrams for  $hh \to hh$ amplitudes
 are shown in
Fig.~\ref{fig:proc-hhhh}, and 
the renormalization conditions read
\begin{equation}
\begin{aligned}
 &\d b_H= \Big[\frac{1}{2} \left( \frac{\mu_3}{v} \l -40a_H b_H+84 a_H^3\r
		-13 \lambda a_H^2+3 \lambda b_H\right)\\
   &\hspace{4.3cm} + \left(87 a_H^2 b_H-120 a_H^4-7 b_H^2\right)\frac{m_h^2 }{v^2}\Big]\div,\\
  &\d b_{\Box H}= \frac{1}{4} \left[-3 \left(4 a_H^4+b_C^2\right)+30 a_C^2 b_C+10 a_H^2 b_H-36
    a_C^4-b_H^2\right] \div,\\
  &\d a_{\Delta H}= -\frac{3}{4} \left(-7 a_C^2 b_C+a_H^2 b_H+6 a_C^4-2 a_H^4+b_C^2\right) \div,\\
  &\d c_{DH}= \Big[-\frac{3}{4} \left(a_C^2-b_C\right){}^2-\frac{a_H^4}{4}\Big] \div,\\
  &\d \lam= \Big\{\frac{3}{2 v^2} \left[8 \mu_3^2 \left(6 a_H^2-b_H\right)
    -16 \lambda \mu_3 {v} a_H + \lambda^2 v^2\right] \\
  &\hspace{3cm}- 12 \left(-12 \mu_3 a_H b_H+32 \mu_3 a_H^3-6 \lambda v a_H^2
    +\lambda v b_H\right) \frac{ m_h^2}{v^3}\\
    &\label{eq:lamb4} \hspace{3cm}+ 6 \left(-48 a_H^2 b_H+80 a_H^4+3 b_H^2\right) \frac{  m_h^4}{v^4}\Big\}\div.
\end{aligned}
\end{equation}
\section{ The Renormalization Group Equations}
\label{sec:AppB}
This Appendix provides the expressions for the RGE of all couplings
discussed above, at the order considered in this paper:
\begin{align}
 \mudmu a_C & = a_C \left[\left(5a_H^2-3 b_C-b_H\right) \frac{m_h^2}{v^2}-a_H \frac{\mu_3}{v}\right] \nonumber \\
            & + a_C^2 \left(4 a_H \frac{m_h^2}{v^2}-\frac{\mu_3}{v}\right) 
              + 3 a_C^3 \frac{m_h^2}{v^2}
              + b_C \frac{\mu_3}{v} 
              - 4 a_H b_C \frac{m_h^2}{v^2}\,\,, \\
 \mudmu b_C & = b_C \left[2\left(5 a_C^2+8a_C a_H+17 a_H^2-3 b_H\right) \frac{m_h^2}{v^2}
              +\lambda - 2(2 a_C + 5 a_H) \frac{\mu_3}{v}\right] \nonumber\\
            & - 2 b_C^2 \frac{m_h^2}{v^2} 
              + a_C^2 \left(-\lambda +4 a_C \frac{\mu_3}{v}+8 a_H \frac{\mu_3}{v}\right) \nonumber\\
            & - 4 a_C^2 \left(2 a_C^2+4 a_C a_H+6 a_H^2-b_H\right) \frac{m_h^2}{v^2}\,\,,\\
 \mudmu a_H & = a_H \left[\lambda -\left(a_C^2-b_C+17 b_H\right) \frac{m_h^2}{v^2}\right]
              - 12 a_H^2 \frac{\mu_3}{v}
              + 45 a_H^3 \frac{m_h^2}{v^2}
              + 2 b_H \frac{\mu_3}{v} \,\,, \\
 \mudmu b_H & = b_H \left[2 \left(-a_C^2+97 a_H^2+b_C\right) \frac{m_h^2}{v^2}
                         - 44a_H \frac{\mu_3}{v} + 3 \lambda\right]
              - 18 b_H^2 \frac{m_h^2}{v^2} \nonumber\\
            & + a_H^2 \left(-13 \lambda +84 a_H \frac{\mu_3}{v}\right)
              - 240 a_H^4 \frac{m_h^2}{v^2}\,\,, \\
 \mudmu m_h^2 &= m_h^2 \left(\lambda -10 a_H \frac{\mu_3}{v}\right)
               + \left(22 a_H^2-4 b_H\right) \frac{m_h^4}{v^2}
               + \mu_3^2 \,\,, \\
 \mudmu \mu _3 &= \mu _3\left[\left(87 a_H^2-15 b_H\right)\frac{m_h^2}{v^2}+3\lambda\right] 
                - 15 a_H \frac{\mu _3^2}{v} \nonumber\\
              & - 12 \lambda  a_H \frac{m_h^2}{v}
                - \left(96 a_H^3-36 a_H b_H\right) \frac{m_h^4}{v^3} \,\,, \\
 \mudmu \lambda  &= \lambda  \left[4 \left(41 a_H^2-7 b_H\right) \frac{m_h^2}{v^2}
                                   -52 a_H \frac{\mu_3}{v}\right] 
                  + 3 \lambda ^2
                  + 24 \left(6 a_H^2-b_H\right) \frac{\mu_3^2}{v^2}  \nonumber\\
                & - 96 a_H \left(8 a_H^2-3 b_H\right) \frac{\mu_3 m_h^2}{v^3}
                  + 12 \left(80 a_H^4-48 a_H^2 b_H+3 b_H^2\right) \frac{m_h^4}{v^4}\,\,, 
\\
 \mudmu v^2 & = -2\left(a_C^2-b_C\right) m_h^2 \,\,, \\
 \mudmu c_6 & = -\frac{1}{24} \left[2+2 a_C^4+3 b_C^2
                                    - a_C^2 \left(-8+6 b_C\right)\right] \,\,, \\
 \mudmu c_7 & = - c_7 \left[\left(a_C^2-5 a_H^2-b_C+b_H\right) \frac{m_h^2}{v^2}
                            + a_H \frac{\mu_3}{v}\right] \nonumber\\
            & + \frac{1}{2} \left(-2 a_C-a_C^3+a_C^2 a_H-a_H b_C\right)\,\,, \\
 \mudmu a_7 & = - a_7 \left[2\left(a_C^2-5 a_H^2-b_C+b_H\right) \frac{m_h^2}{v^2}
                            + 2 a_H \frac{\mu_3}{v}\right] \nonumber\\
            & + \frac{1}{4} \left[4 a_C^4+2 a_C a_H b_C+b_C \left(-2+4 a_H^2-b_H\right)+a_C^2 \left(4-4 a_H^2-3 b_C+b_H\right)\right] \,\,, \\
 \mudmu c_8 & = - c_8 \left[2 \left(a_C^2-5 a_H^2-b_C+b_H\right)\frac{m_h^2}{v^2}
                          +2 a_H \frac{\mu_3}{v}\right] \nonumber\\
            & + \frac{2}{3} \left[a_C^4-a_C^3 a_H-2 a_C a_H b_C+b_C^2+a_C^2 \left(a_H^2+b_C\right)\right] \,\,, \\
 \mudmu c_9 & = \frac{a_C^2}{2}\,\,,  \\
 \mudmu a_9 & = - a_9 \left[\left(a_C^2-5 a_H^2-b_C+b_H\right) \frac{m_h^2}{v^2}
                            + a_H \frac{\mu_3}{v}\right] \nonumber\\ 
            & - \frac{1}{2} a_C \left(a_C^2+a_C a_H-b_C\right)\,\,, \\
 \mudmu b_9 & = - b_9 \left[2 \left(a_C^2-5a_H^2-b_C+b_H\right) \frac{m_h^2}{v^2}
              + 2 a_H \frac{\mu_3}{v}\right] \nonumber\\
            & +\frac{1}{2} \left[4 a_C^4+4 a_C^3 a_H-4 a_C a_H b_C+b_C^2
                                 + a_C^2 \left(4 a_H^2-5 b_C-b_H\right)\right] \,\,, \\
 \mudmu c_{10} &= - c_{10} \left[\left(a_C^2-5 a_H^2-b_C+b_H\right) \frac{m_h^2}{v^2}
                                + a_H \frac{\mu_3}{v}\right] 
               + a_C \left(a_C^2-a_C a_H+b_C\right) \,\,, \\
 \mudmu a_{10} & = -a_{10} \left[2 \left(a_C^2-5 a_H^2-b_C+b_H\right)\frac{m_h^2}{v^2}
                                + 2 a_H \frac{\mu_3}{v}\right] \nonumber\\
              & + \frac{1}{2} \left(-4 a_C^4-4 a_C a_H b_C+b_C^2+a_C^2 \left(4 a_H^2+b_C-b_H\right)\right) \,\,, \\
 \mudmu c_{11} & = -\frac{1}{6} \left(a_C^2-1\right)^2 \,\,, \\
 \mudmu c_{20} & = - c_{20}\left[2 \left(a_C^2-5 a_H^2-b_C+b_H\right) \frac{m_h^2}{v^2}
                                +2 a_H \frac{\mu_3}{v}\right] \\ 
              & + \frac{1}{6} \left[2 a_C^4-2 a_C^3 a_H+a_C^2 \left(6+2 a_H^2-b_C\right)
                                    +2 a_C a_H b_C-b_C \left(6+3 a_H^2+b_C\right)\right] \nonumber\,\,, \\
 \mudmu c_{\Box H} & = -c_{\Box H} \left[2 \left(a_C^2-5 a_H^2-b_C+b_H\right) \frac{m_h^2}{v^2}+2 a_H \frac{\mu_3}{v}\right] 
                      + \frac{1}{2} \left(-3 a_C^2-a_H^2\right) \,\,, \\
 \mudmu a_{\Box H} & = - a_{\Box H} \left[3 \left(a_C^2-5 a_H^2-b_C+b_H\right) \frac{m_h^2}{v^2}
                                       + 3a_H \frac{\mu_3}{v}\right] \nonumber\\
                  & + 3 a_C^3 + a_H^3-\frac{3 a_C b_C}{2}-\frac{a_H b_H}{2}\,\,,  \\
 \mudmu b_{\Box H} & = -b_{\Box H} \left[4 \left(a_C^2-5 a_H^2-b_C+b_H\right)\frac{m_h^2}{v^2}
                                       + 4 a_H \frac{\mu_3}{v}\right] \nonumber\\
                  & - 18 a_C^4-6 a_H^4+15 a_C^2 b_C-\frac{3 b_C^2}{2}+5 a_H^2 b_H-\frac{b_H^2}{2} \,\,, \\
 \mudmu c_{\Delta H} & = -c_{\Delta H} \left[3 \left(a_C^2-5 a_H^2-b_C+b_H\right) \frac{m_h^2}{v^2}
                                          + 3a_H \frac{\mu_3}{v}\right]
                      + 3 a_C^3-a_H^3-3 a_C b_C \,\,, \\
 \mudmu a_{\Delta H} & = -a_{\Delta H} \left[4 \left(a_C^2-5a_H^2-b_C+b_H\right) \frac{m_h^2}{v^2}
                                          + 4 a_H \frac{\mu_3}{v}\right] \nonumber\\
                    & - \frac{3}{2} \left(6 a_C^4-2 a_H^4-7 a_C^2 b_C+b_C^2+a_H^2 b_H\right)\,\,,  \\
 \mudmu c_{DH} & = -c_{DH} \left[4 \left(a_C^2-5 a_H^2-b_C+b_H\right)\frac{m_h^2}{v^2}
                                + 4 a_H \frac{\mu_3}{v}\right] \nonumber\\
              & - \frac{1}{2} \left[a_H^4 + 3\left(a_C^2-b_C\right)^2\right]\,\,.
\end{align}

\bibliographystyle{JHEP}
\bibliography{September-03-PM-copy}

\end{document}